\documentclass[a4paper,reqno]{amsart}
\usepackage[all]{xy} 
\usepackage{amssymb} 
\usepackage{hyperref}
\usepackage{eucal}

\numberwithin{equation}{section}

\newtheorem{definition}{Definition}[section]
\newtheorem{lemma}[definition]{Lemma}
\newtheorem{theorem}[definition]{Theorem}
\newtheorem{proposition}[definition]{Proposition}
\newtheorem{corollary}[definition]{Corollary}
\newtheorem{remarkth}[definition]{Remark}

\newenvironment{remark}{\begin{remarkth}\upshape}{\hfill$\diamond$\end{remarkth}}

\renewcommand{\emph}[1]{{\bfseries\itshape{#1}}}

\renewcommand{\d}{\mathrm{d}^\circ}

\makeatletter
\newcommand\prol{\@ifstar{\@proldf}{\@prolpf}} 
\def\@prolpf{\@ifnextchar[{\@prolpf@wrt}{\@prolpf@}}
\def\@prolpf@wrt[#1]#2{\@ifnextchar[{\@prolpf@wrt@at{#1}{#2}}{\@prolpf@wrt@{#1}{#2}}}

\def\@prolpf@wrt@at#1#2[#3]{\prolsymbol^{#1}_{#3}#2}
\def\@prolpf@wrt@#1#2{\prolsymbol^{#1}#2}
\def\@prolpf@#1{\@ifnextchar[{\@prolpf@at{#1}}{\@prolpf@@{#1}}}
\def\@prolpf@at#1[#2]{\prolsymbol_{#2}#1}
\def\@prolpf@@#1{\prolsymbol#1}
\def\@proldf{\@ifnextchar[{\@proldf@wrt}{\@proldf@}}
\def\@proldf@wrt[#1]#2{\@ifnextchar[{\@proldf@wrt@at{#1}{#2}}{\@proldf@wrt@{#1}{#2}}}

\def\@proldf@wrt@at#1#2[#3]{\prolsymbol^{*#1}_{#3}#2}
\def\@proldf@wrt@#1#2{\prolsymbol^{*#1}#2}
\def\@proldf@#1{\@ifnextchar[{\@proldf@at{#1}}{\@proldf@@{#1}}}
\def\@proldf@at#1[#2]{\prolsymbol^*_{#2}#1}
\def\@proldf@@#1{\prolsymbol^*#1}
\def\prolsymbol{\mathcal{L}}
\makeatother

\newcommand{\Real}{\mathbb{R}}
\def\bea{\begin{eqnarray}}
\def\eea{\end{eqnarray}}
\def\beann{\begin{eqnarray*}}
\def\eeann{\end{eqnarray*}}
\def\beasn{\begin{sneqnarray}}
\def\eeasn{\end{sneqnarray}}
\def\ben{\begin{enumerate}}
\def\een{\end{enumerate}}
\def\bit{\begin{itemize}}
\def\eit{\end{itemize}}
\def\proof{ ({\sl Proof\/}) }
\def\derpar#1#2{\ds\frac{\partial{#1}}{\partial{#2}}}
\def\derpars#1#2#3{\ds\frac{\partial^2{#1}}{\partial{#2}{\partial
{#3}}}}

\def\qed{\ifvmode\Realemovelastskip\fi
{\unskip\nobreak\hfil\penalty50\hbox{}\nobreak\hfil \hbox{\vrule
height1.2ex width1.2ex}\parfillskip=0pt \finalhyphendemerits=0
\par\smallskip}}

\def\vf{\mathfrak{X}}

\def\d{{\rm d}}

\def\Real{\mathbb{R}}
\def\rk{\mathbb{R}^k}
\def\r{\mathbb{R}}
\def\tkq{T^1_kQ}

\def\qed{\ifvmode\removelastskip\fi
{\unskip\nobreak\hfil\penalty50\hbox{}\nobreak\hfil \hbox{\vrule
height1.2ex width1.2ex}\parfillskip=0pt \finalhyphendemerits=0
\par\smallskip}}

\newcommand{\ds}{\displaystyle}

\begin{document}

\title{Nonholonomic constraints in $k$-symplectic   Classical Field
Theories}
\author[M. de Le\'on]{M. de Le\'on}
\address{M. de Le\'on:
Instituto de Ciencias Matem\'aticas (CSIC-UAM-UC3M-UCM),  Serrano 123, 28006
Madrid, Spain} \email{mdeleon@imaff.cfmac.csic.es}
\author[D.\ Mart\'{\i}n de Diego]{D. Mart\'{\i}n de Diego}
\address{D.\ Mart\'{\i}n de Diego:
Instituto de Ciencias Matem\'aticas (CSIC-UAM-UC3M-UCM)\\ Serrano 123, 28006
Madrid, Spain} \email{d.martin@imaff.cfmac.csic.es}
\author[M. Salgado]{M. Salgado}
\address{Modesto Salgado:
Departamento de Xeometr\'{\i}a e Topolox\'{\i}a,
Facultade de Matem\'{a}ticas,
    Universidade de Santiago de Compostela,
    15782-Santiago de Compostela, Spain}
\email{modesto@zmat.usc.es}

\author[S. Vilari\~no]{S. Vilari\~no}
\address{Silvia Vilari\~no:
Departamento de Xeometr\'{\i}a e Topolox\'{\i}a,
Facultade de Matem\'{a}ticas,
    Universidade de Santiago de Compostela,
    15782-Santiago de Compostela, Spain}
    \email{svfernan@usc.es}


\begin{abstract}

A $k$-symplectic framework for classical field theories subject to
nonholonomic constraints is presented. If the constrained problem is
regular one can construct a projection operator such that the
solutions of the constrained problem are obtained by projecting the
solutions of the free problem. Symmetries for the nonholonomic
system are introduced and we show that for every such symmetry,
there exist a nonholonomic momentum equation. The proposed formalism
permits to introduce in a simple way many tools of nonholonomic
mechanics to nonholonomic field theories.

\end{abstract}

\maketitle


\tableofcontents



\section{Introduction}

During the past decades, much effort has been devoted to the
differential geometric treatment of mechanical systems subject to
nonholonomic constraints. To a large extent the growing interest in
this field has been stimulated by its close connection to problems
in control theory (see, for instance,
\cite{Bloch,Cortes-2002}). In the literature, one can distinguish mainly two different
approaches in the study of systems subjected to a  nonholonomic constraints.
The first one, commonly called \emph{nonholonomic mechanics}, is
based on the d'Alembert's principle. This principle   specifies from the constraints a subbundle of the
tangent bundle, representing the admissible infinitesimal virtual
displacements. The second one is a constrained variational approach
called {\it vakonomic mechanics} \cite{arno}. As is well know, the dynamical
equations generated by both approaches are in general not
equivalent \cite{CLMM}.

In this paper we will study an extension of nonholonomic mechanics
to classical field theories with external constraints.
Nonholonomically constrained field theories have already been studied in the
literature. The mathematical framework for a nonholonomic field
theory that has been proposed in \cite{BLMS-2002} involves, among
others, a generalization of d'Alembert's principle and of the
so-called Chetaev rule that is commonly used in nonholonomic mechanics
to characterize the bundle of constraint forms representing the
admissible reaction forces. The constrained field equations for
classical field theories, are then derived in a finite-dimensional
multisymplectic setting. In \cite{VCLM-2005} the authors continue
and extend the work described in \cite{BLMS-2002}.

 The
multisymplectic formalism, was developed by Tulczyjews school in
Warsaw (see, for instance, \cite{KT}), and independently by Garc\'{\i}a and
P\'{e}rez-Rend\'{o}n \cite{PLG1,PLG2} and Goldschmidt and Sternberg \cite{GS}. This
approach was revised, between others, by Martin \cite{mar,mar2} and Gotay et al \cite{GIMMSY-mm} and, more recently, by Cantrijn et al \cite{CIL1} (see also \cite{LeMaSa1} and references therein).

An alternative way to derive  certain types of the field equations
is to use the $k$-symplectic formalism. The $k$-symplectic formalism
is the generalization to field theories of the standard symplectic
formalism in Mechanics, which is the geometric framework for
describing autonomous dynamical systems. In this sense, the
$k$-symplectic formalism is used to give a geometric description of
certain kinds of field theories: in a local description, those
theories whose Lagrangian does not depend on the base coordinates,
denoted by $(t^1, \ldots , t^k)$ (in many of the cases defining the space-time
coordinates); that is, the $k$-symplectic formalism is only valid for
Lagrangians $L(q^i, v^i_A )$ and Hamiltonians $H(q^i, p^A_i )$ that
depend on the field coordinates $q^i$ and on the partial derivatives
of the field $v^i_A$, or the corresponding moment $p^A_i$. A more general approach has been given in \cite{MaMeSa} using the $k$-cosymplectic formalism.

G\"{u}nther's paper  \cite{gun} gave a geometric Hamiltonian formalism
for field theories. The crucial device is the introduction of a
vector-valued generalization of a symplectic form, called a
polysymplectic form. One of the advantages of this formalism is that
one only needs the tangent and cotangent bundle of a manifold to
develop it. In \cite{MRS-2004} G\"{u}nther's formalism has been revised
and clarified. It has been shown that the polysymplectic structures
used by G\"{u}nther to develop his formalism could  be replaced by the
$k$-symplectic structures defined by Awane \cite{aw1,aw3}. So this
formalism is also called $k$-symplectic formalism (see also
\cite{mt1,mt2,LeMeOuRoSa,MaMeSa}).

Let us remark here that the polysymplectic formalism developed by
Sardanashvily \cite{Sarda2,Sarda,Sarda1}, based on a vector-valued form defined
on some associated fiber  bundle, is a different description of
classical field theories of first order than the polysymplectic (or
$k$-symplectic) formalism proposed by G\"{u}nther (see also \cite{Kana}
for more details).
 We must also remark that the soldering form on the linear
frames bundle is a polysymplectic form, and its study and
applications to field theory, constitute the $n$-symplectic geometry
developed by L. K. Norris in \cite{McN,No2,No3,No4,No5}.

 The
purpose of this paper is to give a $k$-symplectic setting for
first-order classical field theories subject to nonholonomic
constraints. In the $k$-symplectic setting we will construct, under
an appropriate additional condition, a kind of projection operator
that maps solutions of the free problem into solutions of the
constrained problem. Nonholonomic symmetries are introduced and we
show that for every such symmetry, there exist a nonholonomic
momentum equation which reduces to a conservation law when the
constraints are absent.

We analyze some particular cases, for instance,  the case of a
constraint submanifold $\mathcal{M}$ obtained as $k$-copies of a
distribution $D$ in $Q$ has special interest. In this particular
case, we construct a distribution $H$ on
$T^1_kQ=TQ\oplus\stackrel{k}{\ldots}\oplus TQ$ (i.e. the Whitney sum
of $k$ copies of $TQ$) along $\mathcal{M}$  such that for each
$w_q\in\mathcal{M}$, $\; H_{w_q}$ is a $k$-symplectic subspace of
the $k$-symplectic vector space
$(T_{w_q}(T^1_kQ),\omega_L^1(w_q),\ldots,$ $
\omega_L^k(w_q);V(w_q))$ where $(\omega_L^1,\ldots, \omega_L^k;V)$
is the $k$-symplectic structure obtained from $L$. Thus if we
restrict the $k$-symplectic structure of $T^1_kQ$ to $H$, the
equations of the constrained problem take the usual form for a free
problem at each fibre of $H$. This procedure extends that by Bates
and Sniatycki \cite{BS-1993} for the linear case.

The scheme of the paper is as follows. In Section \ref{form} we
recall some basic elements from the $k$-symplectic approach to
(unconstrained) Lagrangian classical field theories. In Section
\ref{nlcft} we discuss the construction of a nonholonomic model for
first-order Lagrangian Classical field theories with external
constraints and we obtain the corresponding nonholonomic field
equations. Next, in Section \ref{np} we construct, under an
appropriate additional condition, a projection operator which maps
solutions of the free problem into solutions of the constrained
problem. In Section \ref{nme} we derive the nonholonomic momentum
equation.  In Section \ref{pc} we analyze some particular
cases and in Section \ref{nhft} we briefly analyze the Hamiltonian
case. Finally in Section \ref{c} we conclude with some general
comments.

All manifolds are real, paracompact, connected and $C^\infty$. All
 maps are $C^\infty$. Sum over crossed repeated indices is understood.

\section{   $k$-symplectic Lagrangian field theory}\label{form}
\subsection{Geometric elements}
\subsubsection{The tangent bundle of $k^1$-velocities of a
manifold }

Let $\tau_Q : TQ \to Q$ be the tangent bundle of $Q$. Let us denote
by $T^1_kQ$ the Whitney sum $TQ \oplus \stackrel{k}{\dots} \oplus
TQ$ of $k$ copies of $TQ$, with projection $\tau : T^1_kQ \to Q$,
$\tau ({v_1}_\mathbf{q},\ldots , {v_k}_\mathbf{q})=\mathbf{q}$, where ${v_A}_\mathbf{q}\in T_\mathbf{q}Q$, $1\leq A\leq k$.

$T^1_kQ$ can be identified with the manifold $J^1_0(\Real^k,Q)$ of
the  {\it $k^1$-velocities    of $Q$}, that is,  $1$-jets of maps
$\sigma:\rk\to Q$  with source at $0\in \Real^k$, say
\[
\begin{array}{ccc}
J^1_0(\Real^k,Q) & \equiv & TQ \oplus \stackrel{k}{\dots} \oplus TQ \\
j^1_{0,\mathbf{q}}\sigma & \equiv & ({v_1}_\mathbf{q},\ldots , {v_k}_\mathbf{q})
\end{array}
\]
where $\mathbf{q}=\sigma (0)$,  and ${v_A}_\mathbf{q}=
\sigma_*(0)(\ds\frac{\partial}{\partial t^A}\Big\vert_{0})$.
$T^1_kQ$ is called  {\it the tangent bundle of $k^1$-velocities of
$Q$} (see \cite{mor}).

If $(q^i)$ are local coordinates on $U \subseteq Q$ then the induced
local coordinates $(q^i , v^i)$, $1\leq i \leq n$, on
$TU=\tau_Q^{-1}(U)$ are given by
$$ q^i(v_\mathbf{q})=q^i(\mathbf{q}),\qquad
  v^i(v_\mathbf{q})=v_\mathbf{q}(q^i)  $$
and  the induced local coordinates $(q^i , v_A^i)$, $1\leq i \leq
n,\, 1\leq A \leq k$, on $T^1_kU=\tau^{-1}(U)$ are given by
$$ q^i({v_1}_\mathbf{q},\ldots , {v_k}_\mathbf{q})=q^i(\mathbf{q}),\qquad
  v_A^i({v_1}_\mathbf{q},\ldots , {v_k}_\mathbf{q})={v_A}_\mathbf{q}(q^i) \, .$$

  \paragraph{\bf  A. Vertical lifts of vector fields from $Q$ to $T^1_kQ$}

\begin{definition} For a vector  $X_\mathbf{q}\in T_qQ$, and for
$A=1,\ldots, k$, we define its  {\it vertical $A$-lift} \,
$(X_\mathbf{q})^{V_A}$ as the local vector field   on the fiber
$\tau^{-1}(\mathbf{q})\subset T_k^1Q$ given by
  $$(X_\mathbf{q})^{V_A}(w_\mathbf{q}) = \displaystyle\frac{d}{ds} ({v_1}_\mathbf{q},\ldots ,
  {v_{A-1}}_\mathbf{q},
  {v_{A}}_\mathbf{q}+s  X_\mathbf{q} ,{v_{A+1}}_\mathbf{q} , \ldots,{v_k}_\mathbf{q})\Big\vert_{s=0} $$
\noindent for all
  points $w_\mathbf{q}=({v_1}_\mathbf{q},\ldots , {v_k}_\mathbf{q}) \in \tau^{-1}(\mathbf{q})\subset T^1_kQ$.
\end{definition}

In  local coordinates, if $X_\mathbf{q} = a^i \,\ds\frac{\partial}{\partial
q^i}\Big\vert_{\mathbf{q}}$ then
\begin{equation}\label{xa}
(X_\mathbf{q})^{V_A}(w_\mathbf{q}) =  a^i \displaystyle\frac{\partial}{\partial
v^i_A}\Big\vert_{w_\mathbf{q}}\quad .
\end{equation}

If $X$ is a vector field on $Q$ then we define its vertical $A$-lift
to $T^1_kQ$, $1\leq A \leq k$, as the vector field $X^{V_A}$ given
by
$$
X^{V_A}(w_\mathbf{q})=(X^i(\mathbf{q})\displaystyle\frac{\partial}{\partial q^i}\Big\vert_{\mathbf{q}})^{V_A}(w_\mathbf{q})= X^i(\mathbf{q})\displaystyle\frac{\partial}{\partial
v^i_A}\Big\vert_{w_\mathbf{q}}=( X^i\circ
\tau)(w_\mathbf{q})\displaystyle\frac{\partial}{\partial v^i_A}\Big\vert_{w_\mathbf{q}}\quad ,$$ then
\[
X^{V_A}=( X^i\circ \tau)\displaystyle\frac{\partial}{\partial v^i_A}
\]
where  $X=X^i \derpar{}{q^i}$.

\bigskip

\noindent{\bf   B. Complete lift of vector fields from $Q$ to
$T^1_kQ$.}

\bigskip

Let $\Phi:Q \to Q$ be a differentiable map then the induced map
$T^1_k\Phi:T^1_kQ \to  T^1_kQ$  defined by
$T^1_k\Phi(j^1_0\sigma)=j^1_0(\Phi \circ \sigma)$ is called the
{\it canonical prolongation} of $\Phi$ and, it is also given by
$$  T^1_k\Phi({v_1}_\mathbf{q},\ldots , {v_k}_\mathbf{q})=(\Phi_*(\mathbf{q})({v_1}_\mathbf{q}),\ldots
,\Phi_*(\mathbf{q})({v_k}_\mathbf{q})) \quad ,$$ where ${v_1}_\mathbf{q},\ldots , {v_k}_\mathbf{q}\in
T_{\mathbf q}Q$, $\mathbf{q}\in Q$.

If $Z$ is a vector field on $Q$, with local $1$-parametric group of
transformations $h_s:Q \to Q$ then the local $1$-parametric group of
transformations $T^1_k(h_s):T^1_kQ \to T^1_kQ$ which is the flow of the vector
field $Z^C$ on $T^1_kQ$, called the complete lift of $Z$. Its local
expression is
\begin{equation}\label{locx0}
Z^C=Z^i\displaystyle\frac{\partial}{\partial q^i} \, + \, v^j_A \ds
\frac{\partial Z^k} {\partial q^j}
\displaystyle\frac{\partial}{\partial v^k_A} \; ,
\end{equation}
where $Z=Z^i\ds\frac{\partial}{\partial q^i}$.
\bigskip

\noindent{\bf   C. Canonical $k$-tangent structure.}

\bigskip

The  {\it canonical $k$-tangent structure} on $T^1_kQ$ is the set
$(S^1,\ldots,S^k)$ of tensor fields  of type $(1,1)$ defined by
$${S^A}(w_\mathbf{q})(Z_{w_\mathbf{q}})=
  (\tau_*(w_\mathbf{q})(Z_{w_\mathbf{q}}))^{V_A}(w_\mathbf{q}), \quad \mbox{for all} \, \, Z_{w_\mathbf{q}}\in
T_{w_\mathbf{q}}(T^1_kQ), \quad w_\mathbf{q}
    \in T^1_kQ,$$  for each $A=1, \ldots , k$.

  {}From (\ref{xa})  we have in
local  coordinates
\begin{equation}\label{localJA}
{S^A}=\displaystyle\frac
{\displaystyle\partial}{\displaystyle\partial v^i_A} \otimes dq^i
\end{equation}
  The tensors ${S^A}$ can be regarded as the
$(0,\ldots,0,\stackrel{A}{1} ,0,\ldots,0)$-lift of the identity
tensor on $Q$ to $T^1_kQ$ defined in Morimoto \cite{mor}.

 In the case $k=1$, $S^1$ is the   well-known canonical tangent
structure (also called  vertical endomorphism) of the tangent bundle (see \cite{cra,grif1,klein}).

\bigskip

\noindent{\bf D. Canonical vector fields.}

\bigskip
Let us denote by $\Delta$ the canonical vector field (Liouville
vector field) of the vector bundle $\tau:T^1_kQ \to Q$. This
  vector field $\Delta$  is the
infinitesimal generator of the following flow $$ \psi:\Real \times
T^1_kQ   \longrightarrow T^1_kQ  \quad , \quad
\psi(s,{v_1}_{\mathbf{q}}, \ldots ,{v_k}_{\mathbf{q}})= (e^s
{v_1}_{\mathbf{q}},\ldots , e^s   {v_k}_{\mathbf{q}})\, ,
$$
and in local coordinates it has the form
\[
\Delta = \sum_{i=1}^n\sum_{B=1}^k v^i_B \derpar{}{v_B^i}.
\]

$\Delta$ is a sum of vector fields $\Delta_1+\ldots+\Delta_k$, where
each $\Delta_A$ is the   infinitesimal  generator of the following
flow $\psi^A:\Real \times T^1_kQ   \longrightarrow T^1_kQ$:
\[
\psi^A(s,{v_1}_{\mathbf{q}}, \ldots
,{v_k}_{\mathbf{q}})=({v_1}_{\mathbf{q}},\ldots,v_{{A-1}_\mathbf{q}},
e^s \, {v_A}_{\mathbf{q}},{v_{A+1}}_{\mathbf{q}},
\ldots,{v_k}_{\mathbf{q}})
\]
and in  local coordinates each $\triangle_A$ has the form
\begin{equation}\label{localca}
\Delta_A = \sum_{i=1}^n v^i_A
\displaystyle\frac{\displaystyle\partial}{\displaystyle
\partial v_A^i}\, \quad 1 \leq A \leq k\, .
\end{equation}

\subsubsection{Second-order partial differential equations in
$T^1_kQ$}

$\,$

\bigskip

\noindent{\bf  $k$-vector fields and integral sections.}

\bigskip

Let
$M$ be an arbitrary smooth manifold.
\begin{definition}  \label{kvector} A section
${\bf X} : M \longrightarrow T^1_kM$ of the projection $\tau$ will
be called a {\rm $k$-vector field} on $M$. \end{definition}

 Since
$T^{1}_{k}M$ is  the Whitney sum $TM\oplus \stackrel{k}{\dots}
\oplus TM$ of $k$ copies of $TM$,
  we deduce that to give a $k$-vector field ${\bf X}$ is equivalent to
 give a family of $k$ vector fields $X_{1}, \dots, X_{k}$ on $M$ defined by projection onto each factor. For this reason we will
denote a $k$-vector field by $(X_1, \ldots, X_k)$.

\begin{definition}
\label{integsect} An {\rm integral section}  of the $k$-vector field
${\bf X}=(X_{1}, \dots,X_{k})$, passing through a point $\mathbf{q}\in M$, is
a map $\psi\colon U_0\subset \Real^k \rightarrow M$, defined on some
neighborhood  $U_0$ of $0\in \Real^k$,  such that
$$
\psi(0)=\mathbf{q}, \, \,
\psi_{*}(\mathbf{t})\left(\frac{\partial}{\partial
t^A}\Big\vert_t\right)=X_{A}(\psi (t))
 \quad , \quad \mbox{\rm for $\mathbf{t}\in U_0$, $1\leq A \leq k$}
$$
or, what is equivalent,  $\psi$ satisfies that
$X\circ\psi=\psi^{(1)}$, where  $\psi^{(1)}$ is the first
prolongation of $\psi$  to $T^1_kM$ defined by
$$
\begin{array}{rccl}\label{1prolong}
\psi^{(1)}: & U_0\subset \Real^k & \longrightarrow & T^1_kM
\\\noalign{\medskip}
 & \mathbf{t} & \longrightarrow & \psi^{(1)}(\mathbf{t})=j^1_0\psi_\mathbf{t}\equiv
 \left(\psi_*(\mathbf{t})\left(\derpar{}{t^1}\Big\vert_\mathbf{t}\right),\ldots,
\psi_*(\mathbf{t})\left(\derpar{}{t^k}\Big\vert_\mathbf{t}\right)\right) \, .
 \end{array}
$$
 A $k$-vector field ${\bf X}=(X_1,\ldots , X_k)$ on $M$ is
{\rm integrable} if there is an integral section passing through
every point of $M$.
\end{definition}

In local coordinates, we have
\begin{equation}
\label{localfi11} \psi^{(1)}(t^1, \dots, t^k)=\left( \psi^i (t^1,
\dots, t^k), \frac{\partial\psi^i}{\partial t^A} (t^1, \dots,
t^k)\right), \, 1\leq A\leq k,\,  1\leq i\leq n \, ,
\end{equation}
and $\psi$ is an integral section of $(X_1,\ldots, X_k)$ if and only
if the following equations holds:
\[\derpar{\psi^i}{t^A}= (X_A)^i\circ \psi\, \quad 1\leq A\leq k,\; 1\leq i\leq n\;.\]

 In the $k$-symplectic formalism,  the solutions of the field equations are described as integral
 sections of some $k$-vector fields. Observe that, in case $k=1$, the definition of integral section
  coincides with
the usual definition of integral curve of a vector field.

\bigskip

\noindent{\bf Second-order partial differential equations in
$T^1_kQ$.}

\bigskip

The aim of this subsection is to characterize the integrable
$k$-vector fields on $T^1_kQ$ whose integral sections are
first prolongations $\phi^{(1)}$ of maps  $\phi:\Real^k\to Q$.

\begin{definition} \label{sode0}
A $k$-vector field $\mathbf{\xi}=(\xi_1,\ldots,\xi_k)$ on $T^1_kQ$,
is called  a {\rm second order partial diffe\-rential equation ({\sc
sopde})} if it is a section of the vector bundle
$T^1_k\tau:T^1_k(T^1_kQ)\rightarrow T^1_kQ$; that is,
$$
T^1_k\tau\circ(\xi_1,\ldots,\xi_k)=Id_{T^1_kQ} \quad ,$$ or
equivalently
$$
\tau_*(w_\mathbf{q})(\xi_A(w_\mathbf{q}))= v_{A_\mathbf{q}}\qquad \makebox{for all $A=1, \ldots
, k$,}
$$
where $w_\mathbf{q}=(v_{1_\mathbf{q}},\ldots, v_{k_\mathbf{q}})\in T^1_kQ$.
\end{definition}

 {\it In the case $k=1$, this is just the definition of a second order
differential equation ({\sc sode}).}

 {}From a direct computation
in local coordinates we obtain that the local expression of a {\sc
sopde} $\mathbf{\xi}=(\xi_1,\ldots,\xi_k) $ is
\begin{equation}\label{localsode1}
\xi_A(q^i,v^i_A)= v^i_A\frac{\displaystyle
\partial} {\displaystyle
\partial q^i}+
(\xi_A)^i_B \frac{\displaystyle\partial} {\displaystyle
\partial v^i_B},\quad 1\leq A \leq k \;,
\end{equation}where $(\xi_A)^i_B$ are functions on $T^1_kQ$.

If $\psi:\Real^k \to T^1_kQ$ is an integral section of
$\mathbf{\xi}=(\xi_1,\ldots,\xi_k)$, locally given by
$\psi(\mathbf{t})=(\psi^i(\mathbf{t}),\psi^i_B(\mathbf{t}))$, then from Definition
\ref{integsect} and (\ref{localsode1}) we deduce
\begin{equation}\label{solsopde}
\frac{\displaystyle\partial\psi^i} {\displaystyle\partial
t^A}\Big\vert_{\mathbf{t}}=\psi^i_A(\mathbf{t})\, ,\qquad
\frac{\displaystyle\partial\psi^i_B} {\displaystyle\partial
t^A}\Big\vert_{\mathbf{t}}=(\xi_A)^i_B(\psi(\mathbf{t}))\, .\end{equation}

 {}From (\ref{localfi11}) and (\ref{solsopde}) we obtain the
following proposition.
\begin{proposition} \label{sope1}
Let $\mathbf{\xi}=(\xi_1,\ldots,\xi_k)$ be an integrable {\sc
sopde}. If $\psi$ is an integral section of ${\bf \xi}$ then
$\psi=\phi^{(1)}$, where $\phi^{(1)}$ is the first prolongation of
the map
$\phi=\tau\circ\psi:\Real^k\stackrel{\psi}{\to}T^1_kQ\stackrel{\tau}{\to}Q$,
and $\phi$ is solution to the system of second order partial
differential equations
\begin{equation}\label{nn1}
\frac{\displaystyle\partial^2 \phi^i} {\displaystyle\partial t^A
\partial t^B       }(\mathbf{t})= (\xi_A)^i_B(\phi^i(\mathbf{t}),
\ds\frac{\partial\phi^i}{\partial t^C }(\mathbf{t}))\quad  1\leq i\leq n\, ;
1\leq A,B\leq k.
\end{equation}
Conversely, if $\phi:\Real^k \to Q$ is any map satisfying
(\ref{nn1}) then $\phi^{(1)}$ is an integral section of
$\mathbf{\xi}=(\xi_1,\ldots,\xi_k)$. \qed
\end{proposition}

 {}From (\ref{nn1}) we deduce that if $\mathbf{\xi}$ is an
integrable {\sc sopde} then $(\xi_A)^i_B=(\xi_B)^i_A$ for all
$A,B=1,\ldots, k$.

The following characterization of  the {\sc sopde}'s, using the
canonical $k$-tangent structure of $T^1_kQ$, can be obtained from
(\ref{localJA}), (\ref{localca}) and (\ref{localsode1}).
\begin{proposition}
    \label{pr235}
A $k$-vector field $\mathbf{\xi}=(\xi_1,\ldots,\xi_k)$ on $T^1_kQ$
is a {\sc sopde}  if, and only if, ${S^A}(\xi_A)=\Delta_A$, for all
$A= 1 \ldots , k$.
\end{proposition}
\proof  This is a direct consequence of the local expressions
(\ref{localJA}), (\ref{localca}) and (\ref{localsode1}) of
$S^A,\,\Delta_A$ and $\xi_A$, respectively. \qed

\subsection{Lagrangian formalism: $k$-symplectic Lagrangian systems}
\protect\label{ksls}

 Consider a Lagrangian function  $L:T^1_kQ  \to \r$.  We now define the  {\it action integral}
 \[{\mathcal J}(\phi)=\ds\int_{U_0} (L\circ \phi^{(1)})(\mathbf{t}) d^k\mathbf{t}\,,\]
 where $d^k\mathbf{t}=dt^1\wedge\ldots\wedge dt^k$ is a volume form on $\rk$, $\phi:U_0\subset\rk\to Q$ is a map, with compact support,
 defined on an open set $U_0$ and $\phi^{(1)}:U_0\subset\rk\to T^1_kQ$
 denotes the first prolongation of $\phi$. A map $\phi$ is called an
 extremal for the above action if
 \[\ds\frac{d}{ds}\mathcal{J}(\tau_s\circ\phi)\Big\vert_{s=0}=0
\]for every flow $\tau_s$ on $Q$ such that $\tau_s(\mathbf{q})=\mathbf{q}$ for all
 $\mathbf{q}$ in the boundary of $\phi(U_0)$. Since such a flow $\tau_s$ is
 generated by a vector field $Z\in\mathfrak{X}(Q)$ vanishing on the
 boundary of $\phi(U_0)$, then we conclude that $\phi$ is an extremal
 if and only if
\[\ds\int_{U_0} \left((\mathcal{L}_{Z^c} L)
\circ \phi^{(1)}\right)(\mathbf{t}) d^k\mathbf{t}=0\,, \] for all
$Z$ satisfying the above conditions, where $Z^c$ is the complete
lift of $Z$ to $T^1_kQ$. Putting $Z=Z^i\derpar{}{q^i}$, taking into
account the expression (\ref{locx0}) for the complete lift $Z^c$ and
integrating by parts we deduce that
$\phi(\mathbf{t})=(\phi^i(\mathbf{t}))$ is an extremal of
$\mathcal{J}$ if and only if
\[\ds\int_{U_0} \left[\displaystyle \sum_{A=1}^k\ds\frac{\partial}{\partial t^A}\Big\vert_{\mathbf{t}}
\left(\frac{\displaystyle\partial L}{\displaystyle
\partial v^i_A}\Big\vert_{\phi^{(1)}(\mathbf{t})} \right)- \frac{\displaystyle \partial
L}{\displaystyle
\partial q^i}\Big\vert_{\phi^{(1)}(\mathbf{t})} \right]Z^i d^k\mathbf{t}=0\,, \]for all values of $Z^i$.
Thus, $\phi$ will be an extremal of $\mathcal{J}$ if and only if

\begin{equation}\label{ELe}
\displaystyle \sum_{A=1}^k\ds\frac{\partial}{\partial
t^A}\Big\vert_{\mathbf{t}} \left(\frac{\displaystyle\partial
L}{\displaystyle
\partial v^i_A}\Big\vert_{\phi^{(1)}(\mathbf{t})} \right)= \frac{\displaystyle \partial
L}{\displaystyle
\partial q^i}\Big\vert_{\phi^{(1)}(\mathbf{t})} \;.
\end{equation} The equations (\ref{ELe}) are called the  {\it
Euler-Lagrange equations} for $L$.

We will now give a geometric version of the above equations.

We introduce a family of $1$-forms  $\theta_L^A$ on $T^1_kQ$, $1\leq
A \leq k$,
  using the $k$-tangent structure, as follows
\[
\theta_L^A=  dL \circ {S^A}  \, \quad 1 \leq A \leq k  \;,
\]
which are locally given by
\[\theta_L^A= \ds\frac{\partial L}{\partial
v^i_A}\, dq^i    \quad .
\]

If we denote by $\omega_L^A=-d\theta_L^A$, in local coordinates we
have
\begin{equation}\label{omegala}\omega_L^A=dq^i \wedge d\left(\ds\frac{\partial
L}{\partial v^i_A}\right)= \ds\frac{\partial ^2 L}{\partial
q^j\partial v^i_A}dq^i\wedge dq^j + \ds\frac{\partial ^2 L}{\partial
v^j_B\partial v^i_A}dq^i\wedge dv^j_B \; .
\end{equation}

\begin{definition} The Lagrangian $L: T^1_kQ\longrightarrow \r $ is
said to be regular if, and only if, the matrix
$\left(\ds\frac{\partial^2 L}{\partial v^i_A \partial v^j_B}\right)$
is not singular.
\end{definition}

\begin{remark}{\rm
Let us observe that $L$ regular if and only if
$(\omega_L^1,\ldots, \omega_L^k)$ is a polysympletic form and
$(\omega_L^1,\ldots, \omega_L^k;V={\rm Ker\,\tau_*})$, is a
$k$-symplectic structure (see \cite{MRS-2004}) (See appendix for the introduction of some basic concepts on $k$-symplectic vector spaces). }
\end{remark}

Since $(T^1_kQ,\omega_L^1,\ldots, \omega_L^k;V)$ is a $k$-symplectic
manifold (see Appendix),  we can define the vector bundle morphism,

\[
\begin{array}{rcl}
 \nonumber \Omega_L^{\sharp}:  T^1_k(T^1_kQ) & \longrightarrow & T^*(T^1_kQ)  \\
  \label{sost}
  {\mathbf{X}}=(X_1,\dots,X_k) & \mapsto &
\Omega_L^{\sharp}(X_1,\dots,X_k) ={\rm
trace}(\imath_{X_B}\omega_L^A)= \displaystyle \sum_{A=1}^k \,
\imath_{X_A}\omega_L^A \, .
\end{array}
\]

We now denote by $\vf^k_L(T^1_kQ)$ the set of $k$-vector fields
${\bf \xi_L}=(\xi^1_L,\dots,\xi^k_L)$ on $T^1_kQ$ which are
solutions to the equation \begin{equation} \label{genericEL}
\Omega_L^{\sharp}(\xi^1_L,\dots,\xi^k_L)=\d E_L\;,
\end{equation}where $E_L=\Delta (L)-L$. The family
$(T^1_kQ,\omega_L^A,E_L)$ is called a $k$-symplectic
  Lagrangian system.
If each $ \xi_L^A$ is locally given by
$$
\xi_L^A  =  ( \xi_L^A)^i \frac{\partial}{\partial  q^i} + (
\xi_L^A)^i_B\frac{\partial}{\partial v^i_B}\;,
$$
then $(\xi^1_L,\dots,\xi^k_L)$ is a solution to (\ref{genericEL})
if, and only if, $( \xi_L^A)^i$ and $( \xi_L^A)^i_B$ satisfy the
system of equations \beann
  \left( \frac{\partial^2 L}{\partial q^i \partial v^j_A} -
  \frac{\partial^2 L}{\partial q^j \partial v^i_A}
\right) \, ( \xi_L^A)^j - \frac{\partial^2 L}{\partial v_A^i
\partial v^j_B} \, ( \xi_L^A)^j_B &=&
v_A^j \frac{\partial^2 L}{\partial q^i
\partial v^j_A} - \frac{\partial  L}{\partial q^i } \, ,
\\
\frac{\partial^2 L}{\partial v^j_B\partial v^i_A} \, ( \xi_L^A)^i
&=& \frac{\partial^2 L}{\partial v^j_B\partial v^i_A} \, v_A^i \quad
. \eeann

If the Lagrangian is regular, the above equations  are equivalent to
the equations
\begin{equation}\label{locel4}
\ds\frac{\partial^2 L}{\partial q^j \partial v^i_A} v^j_A +
\ds\frac{\partial^2 L}{\partial v_A^i
\partial v^j_B} \, ( \xi_L^A)^j_B =   \ds\frac{\partial  L}{\partial q^i
}\;,\quad 1\leq i \leq n, \,  1\leq A \leq k,
\end{equation}
and
\begin{equation}\label{locel3}
( \xi_L^A)^i= v_A^i \; .
\end{equation}

Thus, if $L$ is a regular Lagrangian,  we deduce:
\begin{itemize}
\item   If ${\bf
\xi_L}=(\xi^1_L,\dots,\xi^k_L)$ is solution of (\ref{genericEL}),
then it is a {\sc sopde}, (see (\ref{locel3})). \item There are solutions of  (\ref{genericEL})
in a neighborhood of each point of $T^1_kQ$ and, using a partition
of unity, global solutions to (\ref{genericEL}). \item  Since ${\bf
\xi_L}=(\xi^1_L,\dots,\xi^k_L)$ is a {\sc sopde}, from Proposition
\ref{sope1} we know that if it is integrable then its integral
sections are first prolongation $\phi^{(1)}:\rk\to T^1_kQ$ of maps
$\phi:\rk \to Q$, and from (\ref{locel4}) we deduce that $\phi$ is
solution of the Euler-Lagrange equations (\ref{ELe}).

So, equations (\ref{genericEL}) can be considered as a geometric
version of the Euler-Lagrange field equations.
\item The equation (\ref{genericEL}) in the case
$k=1$ is $\imath_{\xi} \omega_L = dE_L$, that is the dynamical
equation of the Lagrangian formalism in Mechanics.
\end{itemize}

Along this paper the family $(T^1_kQ,\omega_L^A,E_L)$ will be called
a  {\it $k$-symplectic Lagrangian system.}

\section{Nonholonomic Lagrangian Classical field
theory}\label{nlcft}

We now bring constraints into the picture. Suppose we have a
Lagrangian $k$-symplectic system on $T^1_kQ$, with a regular
Lagrangian $L$. Let ${\mathcal M}\hookrightarrow T^1_kQ$ be a
submanifold of $T^1_kQ$ of codimension $m$, representing some
external constraints imposed on the system. Although one can
consider more general situations, for the sake of clarity we will
confine ourselves to the case that
 $\mathcal{M}$ projects onto the whole of $Q$, i.e.
$\tau(\mathcal{M})=Q$ and, the restriction
$\tau\vert_{\mathcal{M}}\colon \mathcal{M}\to Q$ of $\tau$ to
$\mathcal{M}$ is a (not necessarily affine) fibre bundle.

Since $\mathcal{M}$ is a submanifold of $T^1_kQ$, one may always
find a covering $\mathcal{U}$ of $\mathcal{M}$ consisting in open
subsets $U$ of $T^1_kQ$, with $\mathcal{M}\cap U\neq\emptyset$, such
that on each $U\in\mathcal{U}$ there exist $m$ functionally
independent smooth functions $\Phi_\alpha$ that locally determine
$\mathcal{M}$, i.e. \[\mathcal{M}\cap
U=\{w_\mathbf{q}=(v_{1_\mathbf{q}},\ldots,v_{k_\mathbf{q}})\in
T^1_kQ|\;\Phi_\alpha(w_\mathbf{q})=0\quad\makebox{for}\quad 1\leq \alpha\leq
m\}\,.\]

The assumption that $\tau\vert_{\mathcal{M}}$ is a fibre bundle
implies, in particular, that the matrix $(\partial
\Phi_\alpha/\partial v^i_A)(w_\mathbf{q})$ has maximal rank $m$ at each point
$w_\mathbf{q}\in \mathcal{M}\cap U$.

\subsection{The bundle of constraint forms}

 We now introduce a special subbundle $F$ of rank $m$  of the bundle of
$\rk$-valued $1$-forms on $T^1_kQ$ defined along the constraint
submanifold $\mathcal{M}$. The elements $\eta$ of $F$ are
$\rk$-valued $1$-forms defined along $\mathcal{M}$ which  are
semi-basic, i.e. $\eta$ vanishes on the $\tau$-vertical vector
fields.

The bundle $F$ is locally generated by $m$ independent $\rk$-valued
$1$-forms $\eta_\alpha$ that locally read
\begin{equation}\label{ext forze}
\eta_\alpha=(\eta_\alpha^1,\ldots,\eta_\alpha^k)=(\eta^1_{\alpha\,i}dq^i,\ldots,\eta^k_{\alpha\,i}dq^i)\,,\end{equation}
for some smooth functions $\eta^A_{\alpha\,i}$ on
$\mathcal{M}\subset T^1_kQ$. The independence of the forms
$\eta_\alpha$ clearly implies that the $m\times kn$-matrix whose
elements are the functions ${\eta_\alpha^A}_i$, has constant maximal
rank $m$ (let us observe that $m$ is exactly the codimension of
$\mathcal{M}$). $F$ is called the  {\it bundle of constraints
forms}.

\begin{remark}{\rm One interesting case is when  $F$ is determined
by $\mathcal{M}$ through application of a ``Chetaev principle". If
the constraint submanifold is giving by the vanishing of $m$
functionally independent functions $\Phi_\alpha$ on $T^1_kQ$, $\,F$
is generated by the following $\rk$-valued forms:
\[\eta_\alpha=({S^1}^*(d\Phi_\alpha),\ldots,{S^k}^*(d\Phi_\alpha))=(
\ds\frac{\partial \Phi_\alpha}{\partial v^i_1}\,dq^i,\ldots,
\ds\frac{\partial \Phi_\alpha}{\partial v^i_k}\,dq^i)
\,.\]}\end{remark}

\subsection{The constraint distribution}\label{dist}

In the sequel, we will  show that the constraint bundle $F$ gives rise to a
distribution ${\mathcal S}$ along $\mathcal{M}$, called the  {\it constraint
distribution}. As above, take $F$  generated by
$m$ $\rk$-valued $1$-forms $\eta_\alpha=(\eta_\alpha^1,\ldots,
\eta_\alpha^k)$ of the form (\ref{ext forze}).

Firstly we introduce the following vector bundle morphisms

\begin{equation}\label{bemol}
\begin{array}{rcl}
 \nonumber \Omega_L^{\flat}:  T(T^1_kQ) & \longrightarrow & (T^1_k)^*(T^1_kQ)  \\
    X & \mapsto &
\Omega_L^{\flat}(X) =(\imath_{X}\omega_L^1,\ldots,\imath_{X}\omega_L^k)
\, ,
\end{array}
\end{equation}
where for an arbitrary manifold $M$ we denote by
$(T^1_k)^*M$ the Whitney sum $T^*M\oplus\stackrel{k}{\ldots}\oplus
T^*M$ of $k$-copies of $T^*M$.

 For each $\alpha\;,(\alpha=1,\ldots, k)$, let $Z_\alpha\in\vf(T^1_kQ)$ be the unique local
   vector field on $T^1_kQ$ defined by
\begin{equation}\label{zalpha}
\tau_*(Z_{\alpha})=0\quad \makebox{and}\quad
\Omega_L^{\flat}(Z_\alpha)= -\,\eta_\alpha\,.
\end{equation}

If we write $$Z_\alpha=(Z_\alpha)^j \frac{\displaystyle\partial
}{\displaystyle
\partial q^j}+ (Z_\alpha)^j_B \frac{\displaystyle\partial }{\displaystyle
\partial v^j_B}$$
we deduce   from (\ref{omegala}) and (\ref{zalpha}) that
$$
(Z_\alpha)^i=0 \;, \quad (Z_\alpha)^j_B\ds\frac{\partial^2
L}{\partial v^j_B\partial v^i_A}= {\eta^A_\alpha}_i\,,  $$ which
determines the $(Z_\alpha)^j_B$ uniquely, since $L$ is supposed to
be regular.

One obtains that:
\[
Z_\alpha=W^{ij}_{AB}\,{\eta^A_\alpha}_i\ds\frac{\partial}{\partial
v^j_B}\;,\] where $(W^{ij}_{AB})$ denotes the inverse matrix of the
Hessian matrix $\left(\ds\frac{\partial^2 L}{\partial v^i_A\partial
v^j_B}\right)$.

The independence of the vector fields $Z_\alpha$ is consequence of
the independence of the $1$-forms $\eta_\alpha$. Thus the vector
fields $Z_\alpha$ span a $m$-dimensional distribution ${\mathcal
S}$, which we will called  {\it the constraint distribution}. This
distribution $\mathcal{S}$ will be used in Section \ref{np}.

\subsection{The nonholonomic field equations}\label{nh eq}

Summarizing, we are looking for a nonholonomic field
theory built on the following objects:
\begin{enumerate}
\item[(i)] a regular Lagrangian $L$;
\item[(ii)] a constraint submanifold $\mathcal{M}\hookrightarrow
T^1_kQ$ which can be locally represented by equations of the form
$\Phi_\alpha(q^i,v^i_A)=0$ for $\alpha=1,\ldots,m$, where the
matrix $(\partial \Phi_\alpha/\partial v^i_A)$ has maximal rank $m$;
\item[(iii)] a bundle $F$ of constraint forms  and an induced
constraint
 distribution $\mathcal{S}$, both defined along $\mathcal{M}$, where $F$ is generated by
 the $m$ independent  semibasic $\rk$-valued
 $1$-forms (\ref{ext forze}).
\end{enumerate}

To complete our model for nonholonomic field theory, we now have to
specify the field equations. We will now introduce the definition of
a solution of the nonholonomic constrained problem using a
generalization of d'Alembert's principle.

\subsubsection{The d'Alembert's principle}

Proceeding as in the case of unconstrained field theories, we
introduce the following definition:
 \begin{definition} A map $\phi:U_0\subset\rk\to Q$  defined on an open
set $U_0\subset Q$ with compact support, is a solution of the
constrained problem under consideration if
$\phi^{(1)}(U_0)\subset\mathcal{M}$ and \[\ds\int_{U_0}
\left((\mathcal{L}_{Z^c} L) \circ \phi^{(1)}\right)(\mathbf{t}) d^k\mathbf{t}=0\,,\]
for each vector field $Z$ on $Q$ that vanish on the boundary of
$\phi(U_0)$ and such that
\begin{equation}\label{var field} \imath_{Z^c}\eta=0\end{equation}
for all $\eta$ of the bundle $F$ of constraint
forms.\end{definition}

Putting $Z=Z^i\derpar{}{q^i}$ and taking into account the expression
(\ref{locx0}) for the complete lift $Z^c$, it is easily seen that
the condition (\ref{var field}) translates into
\[{\eta_\alpha^A}_iZ^i=0
\,,\quad 1\leq \alpha \leq m\,,\quad 1\leq A\leq k\,,\]where
${\eta_\alpha^A}_i$ are the coefficients of the constraint forms
introduced in (\ref{ext forze}).

One can  verify that  $\phi(\mathbf{t})=(\phi^i(\mathbf{t}))$ is a
solution of the constrained problem if and only if
\[\ds\int_{U_0} \left[\displaystyle \sum_{A=1}^k\ds\frac{\partial}{\partial t^A}\Big\vert_{\mathbf{t}}
\left(\frac{\displaystyle\partial L}{\displaystyle
\partial v^i_A}\Big\vert_{\phi^{(1)}(\mathbf{t})} \right)- \frac{\displaystyle \partial
L}{\displaystyle
\partial q^i}\Big\vert_{\phi^{(1)}(\mathbf{t})} \right]Z^id^k\mathbf{t}=0\,, \]for all values of
$Z^i$ satisfying (\ref{var field}).

Therefore, a solution $\phi$ would satisfy the following  system of
partial differential  equations:
\begin{eqnarray}\label{E-L nh}
 \nonumber  \frac{\displaystyle \partial
L}{\displaystyle
\partial q^i}\Big\vert_{\phi^{(1)}(\mathbf{t})} -\displaystyle \sum_{A=1}^k\ds\frac{\partial}{\partial t^A}\Big\vert_{\mathbf{t}}
\left(\frac{\displaystyle\partial L}{\displaystyle
\partial v^i_A}\Big\vert_{\phi^{(1)}(\mathbf{t})} \right)&=& \lambda^\alpha_A\,
{\eta_\alpha^A}_i(\phi^{(1)}(\mathbf{t}))\quad (i=1,\ldots, n)\,, \\
&&
\\\nonumber
  \Phi_\alpha(\phi^{(1)}(\mathbf{t})) &=& 0\quad (\alpha=1,\ldots, m)\,.
\end{eqnarray} As usual, the (a priori) unknown functions
$\lambda^\alpha_A$ play the role ``Lagrange multipliers". The
equations (\ref{E-L nh}) are called the  {\it nonholonomic
Lagrangian field equations} for the constrained problem (compare with \cite{VCLM-2005}).

\subsubsection{Geometric
description of the nonholonomic Lagrangian field equations.}

Consider the following system of equations:
\begin{equation}\label{el1 gen} \Omega_L^{\sharp}(X_1,\ldots,X_k) - dE_L\in
\big<\eta^B_\alpha\big> \;,\quad     X_A\in T\mathcal{M}, \ 1\leq
A\leq k\end{equation}along $\mathcal{M}$.

One obtains
\begin{proposition}
Let ${\bf X}=(X_1, \ldots , X_k)$ be an integrable $k$-vector field
solution of (\ref{el1 gen}). We have
\begin{itemize}
\item[(i)] ${\bf X}=(X_1, \ldots , X_k)$ is a {\sc sopde}.

\item[(ii)] If $\,\phi^{(1)}=(\phi^i(\mathbf{t}),\partial \phi^i/\partial t^A)$ is
an integral section of $\,{\bf X}$ then $\phi$ is solution of the
nonholonomic Lagrangian field equations (\ref{E-L nh}).\end{itemize}
\end{proposition}
\proof Let ${\bf X}=(X_1, \ldots , X_k)$ be an integrable $k$-vector
field solution of (\ref{el1 gen}).  Taking into account that
\(\eta^B_\alpha={\eta^B_\alpha}_idq^i \,,\) (see (\ref{ext forze}))
one obtains that the equation (\ref{el1 gen}) can be written as
follow:
\begin{equation} \label{genericEL*}
\sum_{A=1}^k \imath_{X_A}\omega_L^A-\d E_L=
\lambda^\alpha_B\;{\eta^B_\alpha}_i
 \, dq^i
\end{equation}
where $E_L=\Delta L-L$.

If each $X_A$ is locally given by
$$
X_A  =  (X_A)^i \frac{\partial}{\partial  q^i} + (X
_A)^i_B\frac{\partial}{\partial v^i_B}\;,
$$
then $( X _1, \ldots , X_k)$ is a solution to (\ref{genericEL*}) if
and only if $( X_A)^i$ and $( X _A)^i_B$ satisfy the system of
equations
$$\begin{array}{l}
  \left( \ds\frac{\partial^2 L}{\partial q^i \partial v^j_A} -
   \ds\frac{\partial^2 L}{\partial q^j \partial v^i_A}
\right) \, (( X _A)^j-v^j_A)\\
\qquad  - \left(\ds\frac{\partial^2 L}{\partial
q^j
\partial v^i_A} \, ( X _A)^j+  \ds\frac{\partial^2 L}{\partial v^j_B
\partial v^i_A} (X_A)^j_B -
   \ds\frac{\partial  L}{\partial q^i }\right) = \lambda^\alpha_B\;{\eta^B_\alpha}_i\, ,
\\ \noalign{\medskip}
\ds\frac{\partial^2 L}{\partial v^j_B\partial v^i_A} ( ( X_A)^i
-v^i_A) =0\,   \; .
\end{array}$$

If the Lagrangian is regular, then $(X_A)^j=v^j_A$, that is $( X _1,
\ldots , X_k)$ is a SOPDE,  and the above equations are equivalent
to the equations
\begin{eqnarray}
\nonumber  \ds\frac{\partial^2 L}{\partial q^j \partial v^i_A} v^j_A
+ \ds\frac{\partial^2 L}{\partial v_A^i
\partial v^j_B} \, (X_A)^j_B -  \ds\frac{\partial  L}{\partial q^i
} &=& - \lambda^\alpha_B\;{\eta^B_\alpha}_i \quad (i=1,\ldots, n)\\
\label{locel4*}   &&   \\
\nonumber  (X_A)^i &=& v^i_A\quad (A=1,\ldots,k)\,.
\end{eqnarray}

We will now prove $(ii)$.

Let $\phi^{(1)}=(\phi^i(\mathbf{t}),\partial \phi^i/\partial t^A)$
be an integral section of $( X _1, \ldots , X_k)$ passing through a
point $w_\mathbf{q}\in \mathcal{M}$, that is,
\begin{equation}\label{pp*}\phi^{(1)}(0)=w_\mathbf{q}\in
\mathcal{M}\;,\quad v^j_A\circ\phi^{(1)}=\ds\frac{\partial
\phi^j}{\partial t^A} \;, \quad
(X_A)^j_B\circ\phi^{(1)}=\ds\frac{\partial^2 \phi^j}{\partial t^A
\partial t^B} \,.\end{equation} Substituting (\ref{pp*}) in the first group of equations
(\ref{locel4*}) we obtain the equations $$\displaystyle
\sum_{A=1}^k\ds\frac{\partial}{\partial t^A}\Big\vert_{\mathbf{t}}
\left(\frac{\displaystyle\partial L}{\displaystyle
\partial v^i_A}\Big\vert_{\phi^{(1)}(\mathbf{t})} \right)- \frac{\displaystyle \partial
L}{\displaystyle
\partial q^i}\Big\vert_{\phi^{(1)}(\mathbf{t})}=
  -\, \lambda^\alpha_B\;{\eta^B_\alpha}_i(\phi^{(1)}(t)) \quad (i=1,\ldots,
n)\;,$$ which are the first group of equations in (\ref{E-L nh}).

Finally, from (\ref{pp*}) and since $X_A\big\vert_\mathcal{M}\in
T\mathcal{M}$  we have
$$0=X_A(\Phi_\alpha)=v^i_A\derpar{\Phi_\alpha}{q^i} +
(X_A)^j_B\derpar{\Phi_\alpha}{v^j_B}=
\derpar{\phi^i}{t^A}\derpar{\Phi_\alpha}{q^i} +
\derpars{\phi^j}{t^A}{t^B}\derpar{\Phi_\alpha}{v^j_B}=\derpar{(\Phi_\alpha\circ\phi^{(1)})}{t^A}\Big\vert_{\mathbf{t}}$$
then $\Phi_\alpha\circ\phi^{(1)}$ is a constant function and since
$\phi^{(1)}(0)\in \mathcal{M}$ one obtains
$\Phi_\alpha(\phi^{(1)}(0))=0$ and thus
$\Phi_\alpha(\phi^{(1)}(\mathbf{t}))=0$, that is, the second group
of the equations in (\ref{E-L nh}). Therefore, we conclude that
$\phi$ is solution to the equation (\ref{E-L nh}).\qed

\section{The nonholonomic projector}\label{np}

The purpose of the present section is to show that  for a
nonholonomic first-order field theory in the sense described above,
one can construct, under an appropriate additional condition, a
projection operator which maps solutions of the equation
(\ref{genericEL}) for the unconstrained Lagrangian problem into
solutions of the nonholonomic equations (\ref{el1 gen}).

Given a  {\it constrained problem} with regular Lagrangian $L$,
constraint manifold $\mathcal{M}\subset T^1_kQ$ and constraint
distribution ${\mathcal S}$, we now impose the following  {\it compatibility
condition}: for each $w_\mathbf{q}\in \mathcal{M}$
\begin{equation}\label{compat}
T_{w_\mathbf{q}}\mathcal{M}\cap {\mathcal
S}(w_\mathbf{q})=\{0\}\,.\end{equation}

If $\mathcal{M}$ is locally given by $m$ equations
$\Phi_\alpha(q^i,v^i_A)=0$ and, if ${\mathcal S}$ is locally generated by the
vector fields $Z_\alpha$ (see Subsection \ref{dist}), a
straightforward computation shows that the compatibility condition
is satisfied if and only if \[\det(Z_\alpha(\Phi_\beta)(w_\mathbf{q}))\neq
0\,,\]
 at each point $w_\mathbf{q}\in\mathcal{M}$. Indeed, take $v\in T_{w_\mathbf{q}}\mathcal{M}\cap
 \mathcal{S}(w_\mathbf{q})$. Then $v=v^\alpha Z_\alpha(w_\mathbf{q})$, for some coefficients
 $v^\alpha$. On the other hand, $0=v(\Phi_\beta)=v^\alpha
 Z_\alpha(\Phi_\beta)(w_\mathbf{q})$. Hence, if the matrix
 $(Z_\alpha(\Phi_\beta)(w_\mathbf{q}))$ is regular, we may conclude that
 $v=0$ and the compatibility condition holds.
 The converse is similar: let us suppose that the compatibility
 condition holds. If the matrix $(Z_\alpha(\Phi_\beta)(w_\mathbf{q}))$ is not
 regular, then there exist some vector $v=v^\alpha Z_\alpha(w_\mathbf{q})\neq 0$
 such that $v(\Phi_\beta)=0$ and thus $v\in T_{w_\mathbf{q}}\mathcal{M}\cap
 \mathcal{S}(w_\mathbf{q})$; therefore we conclude that if the compatibility condition
 holds, then the matrix $(Z_\alpha(\Phi_\beta)(w_\mathbf{q}))$ is regular.

 We now have the following result.

\begin{proposition}
If the  compatibility condition {\rm (\ref{compat})} holds, then at each
point $w_\mathbf{q}\in \mathcal{M}$ we have the decomposition
\[T_{w_\mathbf{q}}(T^1_kQ)=T_{w_\mathbf{q}} \mathcal{M}\oplus {\mathcal S}(w_\mathbf{q})\,.\]
\end{proposition}
\proof The proof immediately follows from (\ref{compat}) and a
simple counting of dimensions:
\begin{eqnarray*}
\dim T_{w_\mathbf{q}} \mathcal{M}\oplus
{\mathcal S}(w_\mathbf{q})&=& \dim T_{w_\mathbf{q}} \mathcal{M} + \dim {\mathcal S}(w_\mathbf{q})\\
&=& (n+nk-m)+m=n+nk=
\dim T_{w_\mathbf{q}}(T^1_kQ)\,.\end{eqnarray*} \qed

We now introduce the following notation: $T_{\mathcal{M}}(T^1_kQ)$
denotes the restriction of $T(T^1_kQ)$ to the submanifold of
$T^1_kQ$, $\mathcal{M}$.

 The direct decomposition of
$T_{\mathcal{M}}(T^1_kQ)$ determines two complementary projection
operators $P$ and $Q$:
 $${ P}:
T_{\mathcal{M}}(T^1_kQ) \to T\mathcal{M}\;, \quad
Q=I-P:T_{\mathcal{M}}(T^1_kQ) \to {\mathcal S}\,,$$ where $I$ is the
identity on $T_{\mathcal{M}}(T^1_kQ)$. The projectors $P$ and $Q$
are respectively written as follows:
\[P=I-\,\mathcal{C}^{\alpha\,\beta}Z_\alpha\otimes
d\Phi_\beta\,,\quad Q=\mathcal{C}^{\alpha\,\beta}Z_\alpha\otimes
d\Phi_\beta\,,\] where $(\mathcal{C}^{\alpha\,\beta})$ is the
inverse of the matrix
$(\mathcal{C}_{\alpha\,\beta}:\,=Z_\alpha(\Phi_\beta))$.

 The direct sum decomposition of $T_{\mathcal{M}}(T^1_k Q)$
determines the following decomposition of $T^1_k(T^1_kQ)$ along
$\mathcal M$:
\[T^1_k(T^1_kQ)=T^1_k\mathcal{M}\oplus{\mathfrak S}\,\]
where for each $w_\mathbf{q}\in\mathcal{M}$, $\;\mathfrak{S}_{w_\mathbf{q}}$ is given
by
\[\mathfrak{S}_{w_\mathbf{q}}={\mathcal S}(w_\mathbf{q})\oplus\stackrel{k}{\ldots}\oplus
{\mathcal S}(w_\mathbf{q})\,.\] If ${\mathcal S}$ is locally generated by the vector fields
$Z_\alpha$, then $$\{(Z_\alpha,0,\ldots,0),\,(0,Z_\alpha,0,\ldots,
0),\,\ldots, (0,\ldots,0,Z_\alpha),\, (\alpha=1,\ldots, m)\}$$ is a local
basis of $\mathfrak{S}$.

The direct sum decomposition of $T^1_k(T^1_kQ)$ along $\mathcal{M}$
determines two complementary projection operators $\mathcal{P}$ and
$\mathcal{Q}$:
$$\mathcal{P}:(T^1_k)_{\mathcal{M}}(T^1_kQ) \to T^1_k\mathcal{M}\;, \quad
\mathcal{Q}:(T^1_k)_{\mathcal{M}}(T^1_kQ) \to \mathfrak{S}\,,$$ given
by $\mathcal{P}({X_1}_{w_\mathbf{q}},\ldots,{X_k}_{w_\mathbf{q}})=(P({X_1}_{w_\mathbf{q}}),
\ldots,P({X_k}_{w_\mathbf{q}}))$ and $\mathcal{Q}=\mathcal{I}-\mathcal{P}$,
where $\mathcal{I}$ is the identity on
$(T^1_k)_{\mathcal{M}}(T^1_kQ)$. Here
$(T^1_k)_{\mathcal{M}}(T^1_kQ)$ denotes the restriction of
$T^1_k(T^1_kQ)$ to the constraint submanifold $\mathcal{M}$.

\begin{proposition}
Let $\xi_L=(\xi_L^1,\ldots,\xi_L^k)$ be a solution of the free
Lagrangian problem, i.e., $\xi_L$ is solution to the equation {\rm
(\ref{genericEL})}, then
$\xi_{L,\,\mathcal{M}}=\mathcal{P}({\xi_L}\Big\vert_{\mathcal{M}})$
is a solution to the constraint Lagrangian problem.
\end{proposition}
\proof By definition of $\mathcal{P}$, we know that
$\mathcal{P}\big({\xi_L}\Big\vert_{\mathcal{M}}\big)=\left(P({\xi_L^1}\Big\vert_{\mathcal{M}}),
\ldots,P({\xi_L^k}\Big\vert_{\mathcal{M}})\right)$ with ${
P}({\xi_L^A}\Big\vert_{\mathcal{M}})\in T\mathcal{M}$. Therefore
$\left({ P}({\xi_L^1}\Big\vert_{\mathcal{M}}),\ldots,{
P}({\xi_L^k}\Big\vert_{\mathcal{M}})\right)$ is a solution to
(\ref{el1 gen}) if and only if $\Omega_L^{\sharp}\left({
P}({\xi_L^1}\Big\vert_{\mathcal{M}}),\ldots,{
P}({\xi_L^k}\Big\vert_{\mathcal{M}})\right)- \d E_L\in
\langle\eta_\alpha^A\rangle$.

We have
\begin{eqnarray*}
&&\Omega_L^{\sharp}\left({
P}({\xi_L^1}\Big\vert_{\mathcal{M}}),\ldots,{
P}({\xi_L^k}\Big\vert_{\mathcal{M}})\right)- \d E_L
\\&&=\Omega_L^{\sharp}\left({\xi_L^1}\Big\vert_{\mathcal{M}} -{
Q}({\xi_L^1}\Big\vert_{\mathcal{M}}),\ldots,{\xi_L^k}\Big\vert_{\mathcal{M}}
-{ Q}({\xi_L^k}\Big\vert_{\mathcal{M}})\right)-dE_L \\
&&=- \ds\sum_{A=1}^k i_{\ds \lambda^\alpha_A \,
Z_\alpha}\omega_L^A =- \ds\sum_{A=1}^k\lambda^\alpha_A  i_{\ds
Z_\alpha}\omega_L^A= \ds\sum_{A=1}^k \lambda^\alpha_A
\eta^A_\alpha\in \langle\eta_\alpha^A\rangle\,.
\end{eqnarray*}

Thus we conclude that
$\xi_{L,M}=\left({P}({\xi_L^1}\Big\vert_{\mathcal{M}}),\ldots,{\mathcal
P}({\xi_L^k}\Big\vert_{\mathcal{M}})\right)$ is a solution to
(\ref{el1 gen}).\qed

\begin{remark}{\rm In the particular case $k=1$ we recover the results in  \cite{LMM-1997}}\end{remark}

\section{The nonholonomic momentum equation}\label{nme}
In this section, we derive the nonholonomic momentum equation, the
nonholonomic counterpart to the well-know Noether theorem. More
precisely, we prove that  for every  {\it nonholonomic Lagrangian
symmetry} there exists a certain partial differential equation which is satisfied by the solutions of the constrained problem,
 reducing to a conservation law when the constraints are absent.

Let $G$ be a Lie group and $\mathfrak{g}$ its Lie algebra. Consider
an action $\Phi\colon G\times Q\to Q$. The Lie group $G$ acts on
$T^1_kQ$ by prolongation of $\Phi$, i.e.
\[T^1_k\Phi_g(v_{1_\mathbf{q}},\ldots,
v_{k_\mathbf{q}})=((T_\mathbf{q}\Phi_g)(v_{1_\mathbf{q}}),\ldots,
(T_\mathbf{q}\Phi_g)(v_{k_\mathbf{q}}))\,.\]

\begin{definition}$\,$\newline\vspace{-0.5cm}
\begin{enumerate}
\item We say that the Lagrangian $L$ is {\rm invariant} under the
group action if $L$ is invariant under the induced action of $G$ on
$T^1_kQ$.
\item We say that the Lagrangian $L$ is {\rm infinitesimally
invariant} if for any Lie algebra element $\xi\in \mathfrak{g}$ we
have $\xi_Q^C(L)=0$, where for  a vector field $X$ on $Q$, $\,X^C$
denotes the complete lift of $X$ of $Q$ to $T^1_kQ$  and $\xi_Q$ is
the fundamental vector field defined
by\[\xi_Q(\mathbf{q})=\ds\frac{d}{ds}\,\Phi(exp(s\xi),\mathbf{q})\big\vert_{s=0}\,\quad
\mathbf{q}\in Q\;.\] When $\xi_Q^C(L)=0$, then $\xi_Q$ will be
called an {\rm infinitesimal Lagrangian symmetry}.
 \end{enumerate}
\end{definition}

Let us now assume that $G$ leaves invariant $L,\,\mathcal{M}$ and
$F$:
\[L\circ T^1_k\Phi_g=L,\quad
T^1_k\Phi_g(\mathcal{M})\subset\mathcal{M}\quad \makebox{and}\quad
(T^1_k\Phi_g)^*(F)\subset F\]
for all $g\in G$.

We consider the vector bundle $\mathfrak{g}^F$ over $Q$, defined as
follows: denote by $\mathfrak{g}^F(\mathbf{q})$ the linear subspace of
$\mathfrak{g}$ consisting of all $\xi\in\mathfrak{g}$ such that
\[\xi_Q^C(w_\mathbf{q})\rfloor F=0\quad \makebox{for all}\;
w_\mathbf{q}\in\mathcal{M}\cap\tau^{-1}(\mathbf{q})\,.\]

We assume that the disjoint union of all $\mathfrak{g}^F(\mathbf{q})$, for
all $\mathbf{q} \in Q$ can be given the structure of a vector bundle
$\mathfrak{g}^F$ over $Q$.

To any section $\widetilde{\xi}$ of $\mathfrak{g}^F$, one can
associate a vector field $\widetilde{\xi}_Q$ on $Q$ according to the
following definition:
\begin{equation}\label{fund vf}
\widetilde{\xi}_Q(\mathbf{q}):\,=[\widetilde{\xi}(\mathbf{q})]_Q(\mathbf{q})\;.\end{equation}

\begin{definition} For each $A$, the  {\it $A^{th}$-component of the nonholonomic momentum map}
$(J^{nh})^A$ is the map $(J^{nh})^A:\mathcal{M}\to
(Sec(\mathfrak{g}^F))^*$ constructed as follows: let
$\widetilde{\xi}$ be any section of $\mathfrak{g}^F$, then we define
$(J^{nh})^A_{\widetilde{\xi}}$ along $\mathcal{M}$ as
\begin{equation}\label{mome
map}(J^{nh})^A_{\widetilde{\xi}}=\imath_{\widetilde{\xi}_Q^C}\theta_L^A\,,\end{equation}
where $\widetilde{\xi}_Q$ is the vector field associated to
$\widetilde{\xi}$ according to {\rm (\ref{fund vf})}. \end{definition}

\begin{remark}{\rm  In the particular case $k=1$, corresponding to the Classical Mechanics,
the above definition coincides with the definition of nonholonomic
momentum map introduced by Marsden   {\it et al} in
\cite{BlKrMaMu-1996}.}\end{remark}

\begin{remark}{\rm The map $(J^{nh})^A_\xi$ is the nonholonomic
 version of the $A^{th}$-component $\widehat{J}(0,\ldots, \stackrel{A}{\xi},\ldots, 0) = \theta_L^A(\xi_{T^1_kQ})$, of the momentum map on the
 polysymplectic manifolds $T^1_kQ$ defined in \cite{MRS-2004} when we
 consider the polysymplectic structure given by
 $\omega_L^A=-d\theta_L^A\,, 1\leq A\leq k$.
}\end{remark}

The relevant role of the nonholonomic momentum map lies in the
nonholonomic momentum equation.
\begin{definition}$\,$\newline\vspace{-0.5cm}
\begin{enumerate}
\item A {\rm nonholonomic Lagrangian symmetry} is a section
$\widetilde{\xi}$ of $\mathfrak{g}^F$ such that
$\widetilde{\xi}_Q^C(L)=0$.
\item A {\rm horizontal nonholonomic symmetry} is a constant section of
$\mathfrak{g}^F$.\end{enumerate}
\end{definition}

\begin{theorem}\label{momentum ec} If $\phi\colon U_0\subset\rk\to Q$ is a
solution of the nonholonomic field equations, then for any
nonholonomic Lagrangian symmetry $\;\widetilde{\xi}$ the associated
components of the momentum map
$(J^{nh})^A_{\widetilde{\xi}}\;(A=1,\ldots,k)$ satisfies the
following {\rm nonholonomic momentum equation}:

\[\ds\sum_{A=1}^k\ds\frac{d}{dt^A}
\left((J^{nh})^A_{\widetilde{\xi}(\phi(\mathbf{t}))}\right)=\ds\sum_{A=1}^k(J^{nh})^A_{\frac{d}{dt^A}\,\widetilde{\xi}(\phi(\mathbf{t}))}\,.\]
along $\mathcal{M}$.
\end{theorem}
\proof
\[0=\widetilde{\xi}_Q^C(L)=\widetilde{\xi}_Q^i\derpar{L}{q^i} +
v^j_A\derpar{\widetilde{\xi}_Q^i}{q^j}\derpar{L}{v^i_A}\,.\]

{}From (\ref{E-L nh}) and taking into account that
$\widetilde{\xi}_{Q}(\phi(\mathbf{t}))\in \mathfrak{g}^F$ one obtains that
the above identity is equivalent to
\begin{equation}\label{nonh eq
1}0=\widetilde{\xi}^i_Q\ds\frac{d}{dt^A}\left(\derpar{L}{v^i_A}\right)
+ v^j_A\derpar{\widetilde{\xi}_Q^i}{q^j}\derpar{L}{v^i_A}=
\ds\frac{d}{dt^A}\left(\widetilde{\xi}^i_Q\derpar{L}{v^i_A}\right) -
\left(\ds\frac{d}{dt^A}\;\widetilde{\xi}\right)^i_Q\derpar{L}{v^i_A}\;,\end{equation}
where the latter equality is consequence of
\[\begin{array}{lcl}&&\ds\frac{d}{dt^A}(\widetilde{\xi}_Q^i)=\ds\frac{\partial}{\partial t^A}(\widetilde{\xi}_Q^i(\phi(\mathbf{t})))=\ds\frac{\partial}{\partial t^A}
\left(\ds\frac{d}{ds}\Big\vert_{s=0}{\rm
exp}(s\widetilde{\xi}(\phi(\mathbf{t})))\cdot\phi(\mathbf{t})\right)
\\\noalign{\medskip}&=&\ds\frac{d}{ds}\Big\vert_{s=0}{\rm
exp}(s\ds\frac{\partial}{\partial
{t}^A}\widetilde{\xi}(\phi(\mathbf{t})))\cdot\phi(\mathbf{t}) +
\derpar{\widetilde{\xi}_Q^i}{q^j}v^j_A=
\derpar{\widetilde{\xi}_Q^i}{q^j}v^j_A +
\left(\ds\frac{d}{dt^A}\widetilde{\xi}\right)^i_Q\;.\end{array}\]

Finally, from (\ref{mome map}) and (\ref{nonh eq 1}) one obtains
\[\ds\sum_{A=1}^k\ds\frac{d}{dt^A}
\left((J^{nh})^A_{\widetilde{\xi}(\phi(\mathbf{t}))}\right)=\ds\sum_{A=1}^k(J^{nh})^A_{\frac{d}{dt^A}\,\widetilde{\xi}(\phi(\mathbf{t}))}\,.\]\qed

\begin{corollary}
If $\widetilde{\xi}$ is a horizontal nonholonomic symmetry, then the
following conservation laws holds:
\[\ds\sum_{A=1}^k\ds\frac{d}{dt^A}
\left((J^{nh})^A_{\widetilde{\xi}(\phi(\mathbf{t}))}\right)= 0\,.\]
\end{corollary}

{ \begin{remark}If we rewrite this section in the particular case
$k=1$ we reobtain the Section 4.2 in \cite{BlKrMaMu-1996}.
\end{remark}
\section{Particular cases}\label{pc}

\noindent{\bf 6.1.   Holonomic constraints.}

A distribution $D$ on $Q$  of codimension $m$ induces an submanifold
$\mathcal{M}\hookrightarrow T^1_kQ$ defined as follows:
$({v_1}_\mathbf{q},\ldots,{v_k}_\mathbf{q})$ is an element of
$\mathcal{M}$ if ${v_A}_\mathbf{q}\in D(\mathbf{q})$ for each
$A\,(A=1,\ldots, k)$. In coordinates, if the annihilator $D^0$ is
spanned by the $1$-forms
$\varphi_\alpha={\varphi_\alpha}_idq^i\,(\alpha=1,\ldots, m)$, then
$\mathcal{M}$ is the set of solutions to the $mk$ equations
$\Phi_\alpha^A={\varphi_\alpha}_iv^i_A=0$.

If $D$ is integrable, the constraints induced by $D$ are said to be
 {\it holonomic}: in this case, $\phi^{(1)}$ takes values in
$\mathcal{M}$ if and only if $\phi$ takes values in a fixed leaf of
the foliation induced by $D$, and we conclude that the constraints
can be integrated to constraints on $Q$.

\noindent{\bf 6.2.    Linear constraints induced by  distributions
on $Q$. }

\noindent {\it The constrained problem:} Let $D_1,\ldots,D_k$ be $k$
distributions on $Q$ and we consider the constraint submanifold
$\mathcal{M}=D_1\oplus\ldots\oplus D_k$ of $T^1_kQ$

If we will assume, for each $A\; (A=1,\ldots, k)$, that $\;D_A$ is
defined by the vanishing of $m_A$ functionally independent functions
$\varphi_{\alpha_A}$ on $Q$, then proceeding as above we obtain that
the constraint submanifold is given by the vanishing of
$m=m_1+\ldots+m_k$ independent functions $\Phi_{\alpha_A}^A$ where
\[\Phi_{\alpha_A}^A(v_{1_\mathbf{q}},\ldots, v_{k_\mathbf{q}})=
\tau^*\varphi_{\alpha_A}(v_{A_\mathbf{q}})=(\varphi_{\alpha_A})_iv^i_A\,.\]

For the bundle $F$ of constraints forms we take the bundle along
$\mathcal{M}$, generated by the $m$ $\,\rk$-valued $\,1$-forms
\[{\eta^A_{\alpha_A}}=(
{S^1}^*(d\Phi_{\alpha_A}^A),\ldots,{S^k}^*(d\Phi_{\alpha_A}^A))=
(0,\stackrel{A}{\ldots,\tau^*\varphi_{\alpha_A}},\ldots,0)\;.\]

\noindent {\it Nonholonomic field equations:} In this particular
case, a straightforward computation shows that the equations
(\ref{E-L nh}) become:
\[
\left\{
\begin{array}{rcl}
 \displaystyle \nonumber \displaystyle \sum_{A=1}^k\ds\frac{\partial}{\partial t^A}\Big\vert_{\mathbf{t}}
\left(\frac{\displaystyle\partial L}{\displaystyle
\partial v^i_A}\Big\vert_{\phi^{(1)}(\mathbf{t})} \right)- \frac{\displaystyle \partial
L}{\displaystyle
\partial q^i}\Big\vert_{\phi^{(1)}(\mathbf{t})} &=& \lambda^{\alpha_A}\,
(\varphi_{\alpha_A})_i(\phi(t))\quad (i=1,\ldots, n)\,, \\
&&
\\\nonumber
  \Phi^A_{\alpha_A}(\phi^{(1)}(\mathbf{t})) &=& 0\quad
  (\alpha_A=1,\ldots, m_A\;,A=1,\ldots, k)\,.
\end{array}
\right.
\]

\noindent{\it The constraint submanifold
$\mathcal{M}=D\oplus\stackrel{k}{\ldots}\oplus D$.}
 Let $D$ be a distribution on $Q$. The particular case
$D_1=\ldots=D_k=D$ has special interest. As above, if we assume that
$D$ to be defined by the vanishing of $m$ functionally independent
functions $\varphi_\alpha$ on $Q$, then the constraint submanifold
$\mathcal{M}=D\oplus\stackrel{k}{\ldots}\oplus D$ is given by the
constraint functions
\[\Phi_\alpha^A(v_{1_\mathbf{q}},\ldots,v_{k_\mathbf{q}})=(\varphi_\alpha)_i\;v^i_A=0\,.\]

We will denote by $D^v$ the distribution on $T^1_kQ$ defined by
$(D^v)^0=\langle\tau^*\varphi_\alpha\rangle$ (see
\cite{LM-1996,LMM-1997} for the case $k=1$). Next, we will prove the
following two results (see Appendix for technical definitions).
\begin{lemma}\label{coisot}
$D^v_{w_\mathbf{q}}$ is $k$-coisotropic in
$(T_{w_\mathbf{q}}(T^1_kQ),\omega_L^1(w_\mathbf{q}),\ldots, \omega_L^k(w_\mathbf{q}),V(w_\mathbf{q}))$
for all $w_\mathbf{q}\in\mathcal{M}$, i.e. $(D^v)^\perp_{w_\mathbf{q}}\subset
D^v_{w_\mathbf{q}}$.\end{lemma}
 \proof In fact, since
$(D^v)^0$ is locally generated by semi-basic $1$-forms, we deduce
that
\[(D^v)^\perp_{w_\mathbf{q}}=S(w_\mathbf{q})\subset V_{w_\mathbf{q}}(T^1_kQ)\subset
D^v_{w_\mathbf{q}}\,\] for all $w_\mathbf{q}\in\mathcal{M}$, where
$(D^v)^\perp_{w_\mathbf{q}}=\{U_{w_\mathbf{q}}\in
T^1_kQ:\omega_L^A(U_{w_\mathbf{q}},W_{w_\mathbf{q}})=0,\quad\forall\quad
W_{w_\mathbf{q}}\in D^v_{w_\mathbf{q}}\}$ denotes the $k$-symplectic
orthogonal   of $D^v_{w_\mathbf{q}}$ and $V_{w_\mathbf{q}}(T^1_kQ)$
the vertical distribution of $T^1_kQ$ at the point
$w_\mathbf{q}$.\qed
\begin{proposition}
The following properties are equivalent:
\begin{enumerate}
\item The compatibility condition holds, that is,
$T\mathcal{M}\cap \mathcal{S}=\{0\}$.
\item The distribution $H=TM\cap D^v$ along $\mathcal{M}$ is $k$-symplectic  in the $k$-symplectic vector bundle
$(T(T^1_kQ),\omega_L^1,\ldots, \omega_L^k,V)$
\end{enumerate}
\end{proposition}
\proof If $T\mathcal{M}\cap S=\{0\}$ then
\[T\mathcal{M}\cap \mathcal{S}= T\mathcal{M}\cap(D^v)^\perp=0\]and
\[T_{w_\mathbf{q}}(T^1_kQ)=T_{w_\mathbf{q}}\mathcal{M}\oplus (D^v)^\perp_{w_\mathbf{q}}\,\quad
\forall w_\mathbf{q}\in\mathcal{M}\,.\]

Hence from Lemma \ref{coisot} we obtain
\[(D^v)_{w_\mathbf{q}} = (T_{w_\mathbf{q}}\mathcal{M}\cap (D^v)_{w_\mathbf{q}})\oplus
(D^v)^\perp_{w_\mathbf{q}} = H_{w_\mathbf{q}} \oplus
(D^v)^\perp_{w_\mathbf{q}}= H_{w_\mathbf{q}} \oplus
\mathcal{S}(w_\mathbf{q})\,.\]

Therefore, from a straightforward computation, we obtains that
$H_{w_\mathbf{q}}\cap H_{w_\mathbf{q}}^\perp=\{0\}$ or, equivalently,
that $H_{w_\mathbf{q}}$ is a $k$-symplectic vector subspace  of
$(T_{w_\mathbf{q}}(T^1_kQ),\omega_L^1({w_\mathbf{q}}),$ $\ldots,
\omega_L^k({w_\mathbf{q}}),V_{w_\mathbf{q}})$.

Conversely, assume that for each ${w_\mathbf{q}}\in\mathcal{M}$,
$\;H_{w_\mathbf{q}}$ is a $k$-symplectic subspace in
$(T_{w_\mathbf{q}}(T^1_kQ),\omega_L^1({w_\mathbf{q}}),\ldots,
\omega_L^k({w_\mathbf{q}}),V_{w_\mathbf{q}})$, that is
$H_{w_\mathbf{q}}\cap H_{w_\mathbf{q}}^\perp=\{0\}$. Take
\[Z\in T_{w_\mathbf{q}}\mathcal{M}\cap \mathcal{S}(w_\mathbf{q})=
T_{w_\mathbf{q}}\mathcal{M}\cap(D^v)^\perp_{w_\mathbf{q}}\subset
T_{w_\mathbf{q}}\mathcal{M}\cap(D^v)_{w_\mathbf{q}}=H_{w_\mathbf{q}}\;.\]
Since $\omega_L^A(w_\mathbf{q})(Z,Y)=0$ for all $A\;(A=1,\ldots, k)$ and for
all $Y\in H_{w_\mathbf{q}}$, we conclude that $Z\in H_{w_\mathbf{q}}^\perp$. Thus
$Z\in H_{w_\mathbf{q}}\cap H_{w_\mathbf{q}}^\perp=\{0\}$ and therefore $Z=0$.\qed

Consider now, the restrictions $\omega_H^A$ and $d_HE_L$ to $H$ of
 $\omega_L^A$ and $d E_L$, respectively. Since $H_{w_\mathbf{q}}$ is
 $k$-symplectic for each $w_\mathbf{q}\in \mathcal{M}$, there exist a
 solution on $H$ of the equation
 \begin{equation}\label{bates eq}
\ds\sum_{A=1}^k \imath_{X_A}\omega_H^A=d_HE_L\,.
 \end{equation}
 The above equation may be considered as the $k$-symplectic  version of
the characterization of nonholonomic mechanics in the case of linear
constraints given by \cite{BS-1993}.

\begin{proposition}If $\xi_{L,\mathcal{M}}$ is a solution to the
constrained problem, i.e.
$\xi_{L,\mathcal{M}}=(\xi_{L,\mathcal{M}}^1,\ldots,
\xi_{L,\mathcal{M}}^k)$ satisfies {\rm (\ref{el1 gen})}, if and only if
$\xi_{L,\mathcal{M}}$ is solution of the equation {\rm (\ref{bates eq})}.
\end{proposition}
\proof $\xi_{L,\mathcal{M}}$ is solution of (\ref{el1 gen}) then
$\xi_{L,\mathcal{M}}$ is a {\sc sopde} and $\xi_{L,\mathcal{M}}^A\in
T_{w_\mathbf{q}}\mathcal{M}$.

On the other hand, since $\xi_{L,\mathcal{M}}$ is a {\sc sopde} we
have

\[\tau^*\varphi_\alpha(\xi_{L,\mathcal{M}}^A)=(\varphi_\alpha)_iv^i_A=0\,,\]
along $\mathcal{M}$, and thus $\xi_{L,\mathcal{M}}^A\in
(D^v)_{w_\mathbf{q}}$. Therefore $\xi_{L,\mathcal{M}}^A\in H$ and
$\xi_{L,\mathcal{M}}$ is trivially a solution of equation
(\ref{bates eq}). Conversely, if $(X_1,\ldots, X_k)$ is solution of
the equation (\ref{bates eq}) then for each $A\; (A=1,\ldots, k)$,
$X_A\in H\subset T\mathcal{M}$ and  it is evident that $(X_1,\ldots,
X_k)$ is solution of (\ref{el1 gen}). \qed

\noindent{\bf 6.3. Linear constraints.}

In the local picture, $L$ is subjected to linear constraints defined
by $m$ local functions $\Phi_\alpha:\tkq \to \r$ of the form
$$\Phi_\alpha(v_{1_\mathbf{q}} ,\ldots
,v_{k_\mathbf{q}})=  \Phi_\alpha(q^i,v^i_A)=\ds\sum_{B=1}^k
(\mu^B_\alpha)_i(\mathbf{q})\,
v^i_B=\ds\sum_{B=1}^k\mu^B_\alpha(v_{B_\mathbf{q}})\,,
$$ where $\mu^B_\alpha$ be $mk$ $1$-forms    on $Q$, $1\leq \alpha
\leq m, \,\, 1\leq B \leq k$, locally given by
$\mu^B_\alpha=(\mu^B_\alpha)_i \, dq^i$.

We consider the constraint submanifold
\[\mathcal{M}=\{w_\mathbf{q}=(v_{1_\mathbf{q}}, \ldots,
v_{k_\mathbf{q}})\in\tkq :  \Phi_\alpha(w_\mathbf{q})=0 \quad \forall \alpha\}\] of
dimension $nk-m$.

We now denote by $D$ the distribution on $Q$ given by
$D^0=\langle\mu^B_\alpha\rangle$\,.
\begin{proposition}
Let $L$ be a  regular Lagrangian and ${\bf X}=(X_1, \ldots , X_k)$
is an integrable $k$-vector field which is a solution of
\begin{equation}\label{el1}  \ds\sum_{A=1}^k i_{X_A}\omega_L^A - dE_L\in
(D^V)^0 \;,\quad     X_A\big\vert_\mathcal{M}\in
T\mathcal{M}\end{equation} where $(D^V)^0=<\tau^*\mu^B_\alpha>$. We
have

$i)$ ${\bf X}=(X_1, \ldots , X_k)$ is a {\sc sopde}.

$ii)$ If $\phi^{(1)}=(\phi^i(\mathbf{t}),\partial \phi^i/\partial t^A)$ is an
integral section of ${\bf X}$, then $\phi$ satisfy the equations
{\rm (\ref{E-L nh})}.
\end{proposition}

\proof Let us observe that $\eta^A_\alpha={S^A}^*(d\Phi_\alpha)=
\derpar{\Phi_\alpha}{v^i_A}dq^i=(\mu^A_\alpha)_idq^i =
\tau^*\mu^A_\alpha$ and thus, in this particular case,
$(D^V)^0=\langle\eta^A_\alpha\rangle.$

Therefore, the equations (\ref{el1}) are equivalent to equations
(\ref{el1 gen}) for the case of linear constraints.\qed

\noindent{\bf 6.4.  Constraints defined by connections}

 Suppose that
$Q$ is a fibred manifold over a manifold $M$, say, $\rho:Q \to M$ is
a surjective submersion. Assume that a connection $\Gamma$ in $\rho:
Q \to M$ is given such that
$$TQ=H\oplus V\rho  \;,$$
where $V\rho=\ker T\rho$.
We take fibred coordinates $(q^a,q^\alpha),\,  \, 1\leq a \leq
n-m,\, \, 1\leq \alpha \leq m,\, \, n=\dim \, Q$. The horizontal
distribution is locally spanned by the local vector fields
$$H_a=(\ds\frac{\partial}{\partial
q^a})^H= \ds\frac{\partial}{\partial
q^a}-\Gamma^\alpha_a(q^b,q^\beta)\ds\frac{\partial}{\partial
q^\alpha} \, ,$$ where  $Y^H$ stands for the horizontal lift to $Q$
of a vector field $Y$ on $M$, and $\Gamma^\alpha_a(q^b,q^\beta)$ are
the Christoffel symbols of $\Gamma$. Thus, we obtain a local
basis of vector fields on $Q$,
$$\{ H_a,V_\alpha=\ds\frac{\partial}{\partial
q^\alpha} \}\; .$$ Its dual basis of $1$-forms is
$$\{\eta_a=dq^a\, , \, \eta_\alpha=\Gamma^\alpha_a \, dq^a+dq^\alpha
\} \, .$$ We deduce that $H^0$ is locally spanned by the $1$-forms
$\{ \eta_\alpha \}$.

In this situation we have
$$\tkq=H\oplus \stackrel{k}{\ldots} \oplus H\oplus V\rho
\stackrel{k}{\ldots} \oplus V\rho\; ,$$ and for each vector $v_A$,
$1\leq A\leq k$  we can write
$$v_A=v^a_A\ds\frac{\partial}{\partial
q^a} +v^\alpha_A\ds\frac{\partial}{\partial q^\alpha}
=v^a_A(\ds\frac{\partial}{\partial q^a}- \Gamma^\alpha_a
\ds\frac{\partial}{\partial q^\alpha})
+(v^a_A\Gamma^\alpha_a+v^\alpha_A) \ds\frac{\partial}{\partial
q^\alpha}=v_A^H+v_A^V$$

We define $$\quad \mathcal{M}=H\oplus \stackrel{k}{\ldots} \oplus
H\subset \tkq\, ;$$  then $w_\mathbf{q}=(v_{1_\mathbf{q}}, \ldots, v_{k_\mathbf{q}})\in
\mathcal{M}$ if and only if $v_{A_\mathbf{q}}\in H$ for all $A=1, \ldots ,
k$,   which means that
$$v_A^\alpha=-v^a_A\Gamma^\alpha_a \quad  \mbox{for all}\quad A=1, \ldots ,
k$$ Thus,
\begin{eqnarray*}
\mathcal{M}&=&\{ w_\mathbf{q}\in \tkq \, : \, v_A^\alpha =-v^a_A\Gamma^\alpha_a, \,1\leq A\leq
k\}\\
&=&\{ w_\mathbf{q}\in \tkq \, : \, \varphi_\alpha (v_{A_\mathbf{q}})=0, \,1\leq A\leq
k \}\, .
\end{eqnarray*}

With the $1$-forms  $\varphi_\alpha=\Gamma^\alpha_a \,
dq^a+dq^\alpha$ on $Q$ we shall consider the $1$-forms
$\tau^*\varphi_\alpha$ on $\tkq$. We now consider the equations
\begin{equation}\label{s1}\begin{array}{l}
\Omega_L^\sharp(X_1,\ldots,X_k) - dE_L\in \langle\tau^*\varphi_\alpha\rangle\\
\noalign{\medskip} X_A\big\vert_{\mathcal{M}}\in T\mathcal{M} \, ,
\quad 1\leq A\leq k \, .
\end{array}
\end{equation}

As in the above example, in this particular situation, these
equations are equivalent to equations (\ref{el1 gen}).

We write the first equation in (\ref{s1}) as follows
$$\ds\ds\sum_{A=1}^k i_{X_A}\omega^A_L - dE_L= \ds\sum_{\alpha=1}^m
\lambda^\alpha \, \tau^*\varphi_\alpha$$ and each $X_A$ as
$$X_A= (X_A)^a \ds\frac{\partial}{\partial q^a}+ (X_A)^\alpha
\ds\frac{\partial}{\partial q^\alpha}+(X_A)^a_B
\ds\frac{\partial}{\partial v^a_B}+X_A)^\alpha_B
\ds\frac{\partial}{\partial v^\alpha_B}\, .
$$

 {}From (\ref{s1}) we deduce the three following identities:
\begin{equation}\label{r1}
v^b_A \ds\frac{\partial^2 L}{\partial q^b\partial v^a_A} + v_A^\beta
\ds\frac{\partial^2 L}{\partial q^\beta\partial v^a_A}+(X_A)^b_B
\ds\frac{\partial^2 L}{\partial v^b_B\partial v^a_A} +(X_A)^\beta_B
\ds\frac{\partial^2 L}{\partial v^\beta_B
\partial v^a_A} -\ds\frac{\partial  L}{\partial q^a}=-\lambda_\beta
\Gamma^{\beta}_a
\end{equation}

\begin{equation}\label{r2}
v^b_A \ds\frac{\partial^2 L}{\partial q^b\partial v^\alpha_A} +
v_A^\beta \ds\frac{\partial^2 L}{\partial q^\beta\partial
v^\alpha_A}+(X_A)^b_B \ds\frac{\partial^2 L}{\partial v^b_B\partial
v^\alpha_A} +(X_A)^\beta_B \ds\frac{\partial^2 L}{\partial v^\beta_B
\partial v^\alpha_A} -\ds\frac{\partial  L}{\partial q^\alpha}=-\lambda_\beta
\Gamma^\beta_\alpha
\end{equation}

\begin{equation}\label{r3}
(X_A)^a=v^a_A \;, \quad (X_A)^\alpha=v^\alpha_A
\end{equation}

If $\psi:U_0\subset \rk \to \tkq$,
$\psi(\mathbf{t})=(\psi^a(\mathbf{t}),\psi^\alpha(\mathbf{t}),\psi^a_A(\mathbf{t}),\psi^\alpha_A(\mathbf{t}))$, is
an integral section of ${\bf X}=(X_1, \ldots , X_k)$, then from
(\ref{r1}), (\ref{r2}) and (\ref{r3}) we deduce that $\psi$ is
solution to the equations
$$\left\{\begin{array}{lcl}
v^a_A(\psi(\mathbf{t}))=\ds\frac{\partial  \psi^a}{\partial
t^A}\Big\vert_{\mathbf{t}} & ; &
v^\alpha_A(\psi(\mathbf{t}))=\ds\frac{\partial \psi^\alpha}{\partial
t^A}\Big\vert_{\mathbf{t}}
\\ \noalign{\medskip}
\ds\sum_{A=1}^k  \ds\frac{\partial}{\partial t^A}\left(
\ds\frac{\partial L}{\partial
v^a_A}\Big\vert_{\psi(\mathbf{t})}\right)& - & \ds\frac{\partial
L}{\partial
q^a}\Big\vert_{\psi(\mathbf{t})}=-\lambda_\alpha\, \Gamma^\alpha_a \\
\noalign{\medskip} \ds\sum_{A=1}^k  \ds\frac{\partial}{\partial
t^A}\left( \ds\frac{\partial L}{\partial
v^\alpha_A}\Big\vert_{\psi(\mathbf{t})}\right)& - &
\ds\frac{\partial L}{\partial
q^\alpha}\Big\vert_{\psi(\mathbf{t})}=-\lambda_\alpha
\end{array}\right.$$
or, in other words,
$$\left\{\begin{array}{l}
v^a_A(\psi(\mathbf{t}))=\ds\frac{\partial  \psi^a}{\partial t^A}, \qquad   v^\alpha_A(\psi(\mathbf{t}))=\ds\frac{\partial
\psi^\alpha}{\partial t^A}
\\ \noalign{\medskip}
\ds\sum_{A=1}^k  \ds\frac{\partial}{\partial t^A}\left(
\ds\frac{\partial L}{\partial
v^a_A}\Big\vert_{\psi(\mathbf{t})}-\Gamma^\alpha_a\ds\frac{\partial
L}{\partial v^\alpha_A}\Big\vert_{\psi(\mathbf{t})}\right) -
\left(\ds\frac{\partial L}{\partial
q^a}\Big\vert_{\psi(\mathbf{t})}- \Gamma^{\alpha}_a
\ds\frac{\partial L}{\partial
q^\alpha}\Big\vert_{\psi(\mathbf{t})}\right)=-\frac{d \Gamma^{\alpha}_A}{dt}\frac{\partial L}{\partial
v^\alpha_A}\Big\vert_{\psi(\mathbf{t})}
\end{array}\right.$$
{ These equations are the nonholonomic Euler-Lagrange equations
(\ref{E-L nh}) for this particular case.}

\noindent{\bf 6.5. The nonholonomic Cosserat rod}

The nonholonomic Cosserat rod is an example of a nonholonomic field
theory studied in \cite{Van-2007} (see also \cite{MV-2008}).
 It describes the motion of a rod which
is constrained to roll without sliding on a horizontal surface.

A Cosserat rod can be visualized as specified by a curve
$s\to\mathbf{r}(t)$ in $\Real^3$, called the {\it centerline}, to
which is attached a frame $
\{\mathbf{d}_1(s),\mathbf{d}_2(s),\mathbf{d}_3(s)\}$ called {\it
director frame}. We consider an inextensible Cosserat rod of lenght
$l$. If we denote the centerline at time $t$ as
$s\to\mathbf{r}(t,s)$, inextensibility allows us to assume that the
parameter $s$ is the arc length. The description of the Cosserat rod
can be see in \cite{Van-2007}.

\noindent {\it The nonholonomic second-order model}. The model
described in \cite{Van-2007} fits into the multisymplectic framework
developed on $J^1\pi$, where we have a  fiber bundle $\pi:Y\to X$,
where usually $X$ plays the role of the space-time and the sections
of this fiber bundle are the fields of the theory. In this
particular case the base space $X$ is $\r\times [0,l]$ (time and
space), with coordinates $(t,s)$ and the total space $Y$ is $X\times
\r^2 \times\mathbb{S}^1$, with fibre coordinates $(x,y,\theta)$. In
this model, the fields are the coordinates of the centerline
$(x(t,s),y(t,s))$ and the torsion angle $\theta(t,s)$

Its Lagrangian is given by
\[\mathcal{L}=\ds\frac{\rho}{2}(\dot{x}^2+\dot{y}^2) +
\ds\frac{\alpha}{2}\dot{\theta}^2 - \ds\frac{1}{2}(\beta(\theta')^2
+ Kk^2)\,,\] where $k=(x'')^2 + (y'')^2$, while the constraints are
given by
\[\dot{x}+R\dot{\theta}y'=0\quad\makebox{and}\quad
\dot{y}-R\dot{\theta}x'=0\;.\] Here $\rho,\,\alpha,\,\beta,\,K$ and
$R$ are real parameters and $\dot{x}=\partial x/\partial
t,\;x'\partial x/\partial s$ (analogous for $y$ and $\theta$). This
model is a mathematical simplification of the real physical problem.

We now modify this model, by a lowering process to obtain a
first-order Lagrangian: we introduce new variables $z=x'$ and $v=y'$
and obtain the modified Lagrangian

\[L=\ds\frac{\rho}{2}(\dot{x}^2+\dot{y}^2) +
\ds\frac{\alpha}{2}\dot{\theta}^2 - \ds\frac{1}{2}(\beta(\theta')^2
+ K((z')^2+(v')^2)) + \lambda(z-x') + \mu(v-y')\,,\] where $\lambda$
and $\mu$ are Lagrange multiplier associated to the constraint
$z=x'$ and $v=y'$. This Lagrangian can be thought as a mapping
defined on $T^1_2Q$ where $Q=\r^2\times \mathbb{S}^1\times
\r^4\equiv\r^7$, and if we rewrite this with the notation introduced
in the section 1 we obtain a $k$-symplectic model where the
Lagrangian $L:T^1_2Q\to \r$ is given by
\[L=\frac{\rho}{2}((v^1_1)^2 +
(v^2_1)^2)+\frac{\alpha}{2}(v^3_1)^2-\frac{\beta}{2}(v^3_2)^2 -
\frac{K}{2}((v^4_2)^2 + (v^5_2)^2) + q^6(q^4-v^1_2) +
q^7(q^5-v^2_2)\,\] subject to constraints
\begin{equation}\label{coser const}v^1_1 + R v^3_1v^2_2=0\quad \makebox{and} \quad
v^2_1-Rv^3_1v^1_2=0\,.\end{equation}

In this case the bundle of reaction forces $F$ is generated by the
following forms:
\[\eta_1=(dq^1 + Rv^2_2dq^3,0)\quad \makebox{and}\quad
\eta_2=(dq^2-Rv^1_2dq^3,0)\,.\]

The nonholonomic fields equations associated to $L$ are given by
\begin{equation}\label{cosserat eq}\left\{
  \begin{array}{rcl}
     \rho\derpars{\phi^1}{t^1}{t^1}\Big\vert_{\mathbf{t}}-\derpar{\phi^6}{t^2}\Big\vert_{\mathbf{t}} &=&\lambda  \\\noalign{\medskip}
     \rho\derpars{\phi^2}{t^1}{t^1}\Big\vert_{\mathbf{t}}-\derpar{\phi^7}{t^2}\Big\vert_{\mathbf{t}} &=&\mu \\\noalign{\medskip}
     \alpha\derpars{\phi^3}{t^1}{t^1}\Big\vert_{\mathbf{t}}-\beta\derpars{\phi^3}{t^2}{t^2}\Big\vert_{\mathbf{t}} &=& R\left(\lambda\derpar{\phi^3}{t^1}\Big\vert_{\mathbf{t}}-\mu\derpar{\phi^3}{t^2}\Big\vert_{\mathbf{t}} \right)\\\noalign{\medskip}
     K\derpars{\phi^4}{t^2}{t^2}\Big\vert_{\mathbf{t}}+\phi^6(\mathbf{t}) &=& 0 \\\noalign{\medskip}
     K\derpars{\phi^5}{t^2}{t^2}\Big\vert_{\mathbf{t}} + \phi^7(\mathbf{t}) &=& 0 \\\noalign{\medskip}
     \phi^4(\mathbf{t})-\derpar{\phi^1}{t^2}\Big\vert_{\mathbf{t}} &=& 0 \\\noalign{\medskip}
     \phi^5(\mathbf{t})-\derpar{\phi^2}{t^2}\Big\vert_{\mathbf{t}} &=& 0\,,
  \end{array}
\right.\end{equation}
  where $\lambda$ and $\mu$ are Lagrange multipliers associated with
  the nonholonomic constraints, $\mathbf{t}=(t^1,t^2)=(t,s)$ are the coordinates time and space and the field
   $\phi:U_0\subset\r^2\to\r^7$  are the coordinates of the centerline
  $(\phi^1(\mathbf{t}),\phi^2(\mathbf{t}))$ and the torsion angle $\phi^3(\mathbf{t})$. As one can see in the equation (\ref{cosserat eq}) the components $\phi^i,\ i\geq 4$ are determined by $(\phi^1,\phi^2,\phi^3)$. These equations are supplemented by
  the constraint equations (\ref{coser const}).

 Consider the action of
  $\r^2\times\mathbb{S}^1$ on $Q$ according to the following
  definition: for each $(a,b,\theta)\in\r^2\times\mathbb{S}^1$ we
  consider the map
  $\Phi_{(a,b,\theta)}(q^1,\ldots,q^7)=(q^1+a,q^2+b,q^3+\theta,q^4,\ldots,q^7)$.

It is easy to see that the following vector field annihilates $F$
along $\mathcal{M}$:\[
\widetilde{\xi}=-Rv^2_2\derpar{}{q^1}+Rv^1_2\derpar{}{q^2}+\derpar{}{q^3}\,.\]
This generalized vector field corresponds with the section
$\widetilde{\xi}=(-Rv^2_2,Rv^1_2,1)$ of $\tau^*\mathfrak{g}^F$.

As $\widetilde{\xi}(L)=0$ the Theorem \ref{momentum ec} can be
applied and the  {\it nonholonomic momentum equation} (see
\cite{MV-2008}) hence becomes
\[R\left(\rho\derpars{\phi^1}{t^1}{t^1} -
\derpar{\phi^6}{t^2}\right)\derpar{\phi^2}{t^2} -
R\left(\rho\derpars{\phi^2}{t^1}{t^1}
-\derpar{\phi^7}{t^2}\right)\derpar{\phi^1}{t^2}=\alpha\derpars{\phi^3}{t^1}{t^1}
- \beta\derpars{\phi^3}{t^2}{t^2}\,.\]

 This nonholonomic conservation law can also be derived from
the nonholonomic equations substituting the two first equation in
(\ref{cosserat eq}) into the three equation. Unfortunately, the
knowledge of this nonholonomic conservation law does not help us in
solving the field equations.

\begin{remark}Rewriting the nonholonomic momentum equation in the
notation used in the description of the second order model, we
obtain the nonholonomic momentum equation for spacial symmetries
given in \cite{MV-2008} into the multisymplectic setting.
\end{remark}

\section{Non-holonomic Hamiltonian field theory}\label{nhft}

We now turn to the Hamiltonian description of the nonholonomic
system on the bundle of $k^1$-covelocities $(T^1_k)^*Q$ of $Q$.

The \emph{Legendre map} $FL\colon T^1_kQ\to (T^1_k)^*Q$ is defined (see \cite{gun,MaMeSa}) as follows: if $(v_{1_\mathbf{q}},\ldots,
v_{k_\mathbf{q}})\in (T^1_k)_\mathbf{q} Q$,
\[[FL(v_{1_\mathbf{q}},\ldots, v_{k_\mathbf{q}})]^A(u_\mathbf{q}) = \frac{d}{ds}\Big\vert_{s=0}L
(v_{1_\mathbf{q}},\ldots, v_{A_\mathbf{q}}+su_\mathbf{q},\ldots, v_{k_\mathbf{q}})\;, \] for each
$A=1,\ldots, k$ and $u_\mathbf{q}\in T_\mathbf{q}Q$. Locally $FL$ is given by
\[
FL(q^i,v^i_A)=(q^i,\ds\derpar{L}{v^i_A})\;.\]

Assuming the regularity of the Lagrangian, we have that the
Lagrangian and Hamiltonian formulations are locally equivalent. If
we suppose that the Lagrangian $L$ is hyperregular, the Legendre
transformation is a global diffeomorphism.

The constraint function on $(T^1_k)^*Q$ becomes
$\Psi_\alpha=\Phi_\alpha \circ FL^{-1}:(T^1_k)^*Q\to \r$, that is,
\[\Psi_\alpha (q^i,p^A_i)=\Phi_\alpha(q^i,\derpar{H}{p^A_i})\,,\]
where the Hamiltonian function $H:(T^1_k)^*Q\to \r$ is defined by
$H=E_L\circ FL^{-1}$. Since locally
$FL^{-1}(q^i,p^A_i)=(q^i,\derpar{H}{p^A_i})$, then
\[H=v^i_A\circ FL^{-1}\;p^A_i-L\circ FL^{-1}\,.\]

Thus, from (\ref{E-L nh}), one obtains
\begin{eqnarray}\nonumber
  \derpar{H}{p^A_i} &=& v^i_A\circ FL^{-1} \\\noalign{\medskip}\nonumber
  \derpar{H}{q^i} &=& -\derpar{L}{q^i}\circ
  FL^{-1}=-\lambda^\alpha_C
  \derpar{\Psi_\alpha}{p^B_k}{\mathcal H}^{ki}_{BC}
  -\ds\sum_{A=1}^k\derpar{}{t^A}\left(\derpar{L}{v^i_A}\circ FL^{-1}\right)\,,
\end{eqnarray}where ${\mathcal H}^{ki}_{BC}$ are the components of the
inverse of the matrix $({\mathcal H}^{BC}_{ik})=(\partial^2 H/\partial
p^B_i\partial p^C_k)$. Note that
\[\derpar{\Psi_\alpha}{p^B_k}{\mathcal
H}^{ki}_{BC}=\derpar{\varphi_\alpha}{v^i_C}\circ FL^{-1}\,.\]

Therefore, the non-holonomic Hamiltonian equations on $(T^1_k)^*Q$
can be written as follows
\[
\left\{
  \begin{array}{lcl}
    \ds\derpar{H}{p^A_i}\Big\vert_{\psi(\mathbf{t})} &=& \derpar{\psi^i}{t^A}\Big\vert_{\mathbf{t}}, \\\noalign{\medskip}
    \derpar{H}{q^i}\Big\vert_{\psi(\mathbf{t})} &=&
    -\lambda^\alpha_C
  \derpar{\Psi_\alpha}{p^B_k}\Big\vert_{\psi(\mathbf{t})}{\mathcal
  H}^{ki}_{BC}(\psi(\mathbf{t}))
  -\ds\sum_{A=1}^k\derpar{\psi^A_i}{t^A}\Big\vert_{\mathbf{t}}\\\noalign{\medskip}
  0 &=&  \Psi_\alpha(\psi^i(\mathbf{t}),\psi^A_i(\mathbf{t}))
  \end{array}
\right. \] where $\psi:U\subset\rk\to (T^1_k)^*Q$ is locally given
by $\psi(\mathbf{t})=(\psi^i(\mathbf{t}),\psi^A_i(\mathbf{t}))$.

Next we will give a geometrical description of these equations.

Let $M\subset (T^1_k)^*Q$ be the image of the constraint submanifold
$\mathcal{M}$ under the Legendre map and let $\mathcal{F}$ the
bundle locally generated by the independent $\rk$-valued $1$-form
\[\widetilde{\eta}_\alpha=FL^*\eta_\alpha,\quad  1\leq \alpha \leq m\,.
\]

Thus, the ``Hamilton equations" for the nonholonomic problem can be
rewritten in intrinsic form as \[ \ds\sum_{A=1}^k i_{X_A}\omega^A -
dH\in \langle\widetilde{\eta}_\alpha^A\rangle\;,\quad
X_A\big\vert_{M}\in T M\,,\] where
$\widetilde{\eta}_\alpha^A=FL^*\eta^A_\alpha$.

\section{Conclusions}\label{c}

We have studied various aspects of  first-order classical field
theories subject to nonholonomic constraints in the $k$-symplectic
framework. The study is very similar to the case  of  particle mechanics
(see [24]) and, also, the results are quite similar to those obtained in the  multisymplectic framework (see \cite{VCLM-2005}) but the $k$-symplectic approach seems simpler in many applications. We
have shown that in the $k$-symplectic approach, the solutions to the
equations for the constrained problem can be obtained by a
projection of the solution to the equations for the unconstrained  Lagrangian
problem.

We analyze the particular case of a constraint submanifold
$\mathcal{M}$ which we obtain as $k$-copies of a distribution on the
configuration space $Q$. In this particular case, we construct a
distribution $H$ on $T^1_kQ$ along $\mathcal{M}$ which is a
$k$-symplectic subspace in $(T(T^1_kQ),\omega_L^1,\ldots,
\omega_L^k;V)$ where $(\omega_L^1,\ldots, \omega_L^k;V)$ is the
$k$-symplectic structure obtained from $L$.
Finally, the nonholonomic momentum map is defined in a similar way than in classical mechanics. The applicability of the theory is shown in some examples and particular cases.

\section*{Appendix: $k$-symplectic vector spaces}

Let $U$ be a  vector space of dimension $n(k+1)$,  $V$ a
subspace of $U$ of codimension $n$ and
$\omega^1,\ldots,\omega^k$, $k$ $2$-forms on $U$. For each $A\;
(A=1,\ldots, k)$,  $\ker\,\omega^A$ denotes the subspace associated
to $\omega^A$ given by
\[\ker\,\omega^A=\{u\in U/ \omega^A(u,v)=0 \quad \forall\; v\in U\}\;.\]

\begin{definition}
$(\omega^1,\ldots, \omega^k;V)$ is a $k$-symplectic structure on $U$
if
\[\omega^A\vert_{V\times V}=0\;,\quad
\ds\bigcap_{A=1}^k  \ker\,\omega^A=0\,.\] We say that
$(U,\omega^1,\ldots, \omega^k;V)$ is a $k$-symplectic vector space.
\end{definition}

Let $W$ be a liner subspace of $U$. The $k$-symplectic orthogonal of
$W$ is the linear subspace of $U$ defined by

\[W^\bot=\{u\in U/\; \omega^A(u,w)=0 \makebox{ for all} \;w\in W,
A=1,\ldots, k\}\,.\]

\begin{proposition}The $k$-symplectic orthogonal satisfies
\begin{enumerate}
\item $A\subset B\Rightarrow B^\bot\subset A^\bot$.
\item $W\subset (W^\bot)^\bot$.
\end{enumerate}\end{proposition}

\begin{remark}{\rm Unlike what happens in the symplectic vector spaces,
in our context in general $\dim W + \dim
 W^\bot \neq\dim U$. In fact, considering for instance, the real
 space $\r^3$ equipped with the $2$-symplectic structure defined by:
 \[\omega^1=e^1\wedge e^3\quad \omega^2=e^2\wedge e^3 \quad V=\ker
 e^3\] where $\{e^1,\, e^2,\,e^3\}$ is the dual basis of the
 canonical basis $\{e_1,\,e_2,\,e_3\}$ of $U=\r^3$. We consider
 $W={\rm span\,}\{e_3\}$, the $2$-symplectic orthogonal of $W$ is
 $W^\bot={\rm span}\{e_3\}$. In this case, $\dim W + \dim W^\bot =
 2\neq \dim \r^3$.}\end{remark}
We can now introduce the following special types of subspaces of a
$k$-symplectic vector space, generalizing the corresponding notions
from symplectic geometry.

\begin{definition}
 Let $(U,\omega^1,\ldots, \omega^k;V)$ be  a $k$-symplectic vector space and $W$ a linear subspace of $U$.
\begin{itemize}
\item $W$ is called {\rm isotropic} if $W\subset W^\bot$.
\item $W$ is {\rm coisotropic} if $W^\bot\subset W$.
\item $W$ is {\rm Lagrangian} if $W=W^\bot$.
\item $W$ is {\rm $k$-symplectic } if $W\cap W^\bot=0$.
\end{itemize}
\end{definition}
\begin{proposition}
For every vector subspace $W$ of $U$ the following properties are
equivalent: \begin{enumerate} \item W is an isotropic subspace.
\item $\omega^A(u,v)=0\; (A=1,\ldots, k)$ for all $u,v\in W$.
\end{enumerate}
\end{proposition}
\proof Let us suppose that $W$ is isotropic, then $W\subset W^\bot$.
Therefore if $u,\,v\in W$ then $\omega^A(u,v)=0$ since $u\in
W^\bot$.

Conversely,  $u\in W$, then for each $v\in W$, $\,\omega^A(u,v)=0\;
(A=1,\ldots, k)$. Therefore $u \in W^\bot$. Thus we can conclude
that $W\subset W^\bot$. \qed

\begin{proposition} Let $(U,\omega^1,\ldots, \omega^k;V)$ be  a $k$-symplectic vector space and $W$ a linear subspace of $U$.
 $W$ is a $k$-symplectic subspace
if and only if $W$ with the restriction of the $k$-symplectic
structure of $U$ to $W$ is a $k$-symplectic vector space. \end{proposition}
\proof Let us suppose that $W$ is a $k$-symplectic subspace of
$(U,\omega^1,\ldots, \omega^k;V)$. Consider the restriction
$\omega^A_W$ to $W$ of the $2$-forms $\omega^A$, we will now prove
that $(W,\omega^1_W,\ldots,\omega^k_W,V\cap W)$ is a $k$-symplectic
vector space.

Given $u,v\in V\cap W$ one obtains \[\omega^A_W(u,v)
=\omega^A\vert_{V\times V}(u,v)=0\; (A=1,\ldots, k)\,.\]

On the other hand, if $u\in \cap\ker \omega^A_W$ then
$\omega^A(u,v)=0 \; (A=1,\ldots, k)$ for all $v\in W$, then $u\in
W^\bot$. Therefore, since $u\in W\cap W^\bot=\{0\}$ we deduce
$\cap\ker \omega^A_W=\{0\}$.

Conversely, if $u\in W\cap W^\bot$ then $u\in W^\bot$, that is
$\omega^A(u,v)=0\; (A=1,\ldots, k)$ for all $v\in W$. Since $u\in W$
we obtain that $u\in\cap\ker \omega^A_W=\{0\}$ an therefore
$u=0$.\qed

\begin{definition}\label{defaw} {\rm (Awane \cite{aw1})}
 A {\rm $k$-symplectic structure} on  a manifold $M$ of dimension $N=n+kn$
is a  family $(\omega^A,V;1\leq A\leq k)$, where each $\omega^A$ is
a closed $2$-form and $V$ is an integrable $nk$-dimensional
distribution on $M$ such that
 $$
(i) \quad \omega^A\vert_{ V\times V}=0,\qquad
 (ii) \quad \cap_{A=1}^{k} \ker\omega^A=\{0\} \ .
$$
Then $(M,\omega^A,V)$ is called a {\rm $k$-symplectic manifold}.
\end{definition}

Let us observe that if $(M,\omega^1,\ldots, \omega^k,V)$ is a
$k$-symplectic manifold, then for each $x\in M$, we have that
$(\omega^1_x,\ldots, \omega^k_x,V_x)$ is a $k$-symplectic structure
on the vector space $T_xM$.

\section*{Acknowledgments}
This work has been partially supported by MEC (Spain) Grant  MTM 2007-62478, project ``Ingenio Mathematica"
(i-MATH) No. CSD 2006-00032 (Consolider-Ingenio 2010) and
S-0505/ESP/0158 of the CAM.
 Silvia Vilari\~{n}o acknowledges the financial support of
 {\it Xunta de Galicia}  Grants IN840C 2006/119-0 and IN809A 2007/151-0.

\end{document}